\documentclass{aa}

\usepackage{graphicx}
\usepackage{txfonts}
\usepackage{natbib}
\usepackage{xcolor}
\usepackage{hyperref}
\usepackage{multirow}

\usepackage{orcidlink}

\hypersetup{
    colorlinks=true,
    linkcolor=blue,
    citecolor=blue,
    urlcolor=blue
}

\begin{document}

\title{Impact of nuclear mass models on $r$-process nucleosynthesis and heavy element abundances in $r$-process enhanced metal-poor stars}

\titlerunning{Impact of nuclear mass models on $r$-process nucleosynthesis}

\author{Meng-Hua Chen$^{\orcidlink{0000-0001-8406-8683}}$
\inst{1}
\and
Li-Xin Li$^{\orcidlink{0000-0002-8466-321X}}$
\inst{1}
\and
En-Wei Liang$^{\orcidlink{0000-0002-7044-733X}}$
\inst{2}
\and
Ning Wang$^{\orcidlink{0000-0002-2629-6730}}$
\inst{3}
}

\authorrunning{M.-H. Chen et al.}

\institute{Kavli Institute for Astronomy and Astrophysics, Peking University, Beijing 100871, China\\
\email{physcmh@pku.edu.cn}
\and
Guangxi Key Laboratory for Relativistic Astrophysics, School of Physical Science and Technology, Guangxi University, Nanning 530004, China\\
\email{lew@gxu.edu.cn}
\and
Guangxi Key Laboratory of Nuclear Physics and Technology, Department of Physics, Guangxi Normal University, Guilin 541004, China
}

\date{Received xxxx; accepted xxxx}

\abstract
{Due to the lack of experimental data on extremely neutron-rich nuclei, theoretical values derived from nuclear physics models are essential for the rapid neutron capture process ($r$-process). Metal-poor stars enriched by the $r$-process offer valuable cases for studying the impact of nuclear physics models on $r$-process nucleosynthesis. This study analyzes four widely used nuclear physics models in detail: Finite-Range Droplet Model, Hartree-Fock-Bogoliubov, Duflo-Zuker, and Weizs$\ddot{\rm a}$cker-Skyrme (WS4). Theoretical values predicted by the WS4 model are found to be in good agreement with experimental data, with deviations significantly smaller than those predicted by other models. The heavy element abundances observed in $r$-process enhanced metal-poor stars can be accurately reproduced by $r$-process nucleosynthesis simulations using the WS4 model, particularly for the rare earth elements. This suggests that nuclear data provided by nuclear physics model like WS4 are both essential and crucial for $r$-process nucleosynthesis studies.}


\maketitle

\section{Introduction}
\label{introduction}

The rapid neutron-capture process ($r$-process) has long been considered the primary mechanism responsible for the production of heavy elements beyond iron in the universe \citep{1957RvMP...29..547B}. The reaction conditions for the $r$-process require high density and temperature, as well as extremely neutron-rich environments. The ejected materials from the merger of binary neutron stars or neutron star-black hole systems are ideal sites for $r$-process nucleosynthesis \citep{1974ApJ...192L.145L,1982ApL....22..143S}. In 1998, \citet{1998ApJ...507L..59L} first predicted that the radioactive decay of freshly synthesized heavy elements via $r$-process nucleosysnthesis could power a bright thermal transient known as a ``kilonova''\citep{2010MNRAS.406.2650M}. This $r$-process kilonova model was confirmed by multi-messenger observations of the first neutron star merger event GW170817/GRB170817A/AT2017gfo \citep{2017ApJ...848L..12A}. The light curve and color evolution of kilonova AT2017gfo suggest that approximately $0.05M_{\odot}$ of heavy $r$-process nuclei were synthesized in this merger event \citep{2017Natur.551...80K,2024MNRAS.527.5540C}. If the heavy element yield in the kilonova AT2017gfo can be regarded as typical for neutron star mergers, it suggests that such merger events could be the dominant contributor to the production of heavy $r$-process elements in the universe \citep{2018IJMPD..2742005H,2024MNRAS.529.1154C}.

The detection of kilonova emission supports the prediction that neutron star mergers are astrophysical sites of $r$-process nucleosynthesis, but determining the detailed composition of heavy elements in the merger ejecta remains a great challenge. \citet{2019Natur.574..497W} analyzed spectral features in the kilonova AT2017gfo and identified an individual heavy element strontium (atomic number $Z=38$). Recently, \citet{2024Natur.626..737L} identified a heavier element tellurium (atomic number $Z=52$) in the kilonova associated with the gamma-ray burst 230307A observed by the James Webb Space Telescope. Another direct approach for identifying heavy elements could be the observation of gamma-ray lines produced by the radioactive decay in the merger ejecta \citep{2016MNRAS.459...35H,2019ApJ...872...19L,2020ApJ...903L...3W,2021ApJ...919...59C,2022ApJ...932L...7C,2024ApJ...971..143C,2024PhRvL.132e2701V}. Unfortunately, current MeV gamma-ray detectors are not sufficient to detect these radioactive gamma-ray lines \citep{2021ApJ...919...59C,2022ApJ...932L...7C,2024ApJ...971..143C}.

The heavy elements ejected by neutron star mergers enrich the interstellar medium and contribute to the formation of next-generation stars \citep{2008ARA&A..46..241S}. Consequently, the detailed composition of these heavy elements has been recorded in some stars enriched by a single $r$-process event, such as $r$-process enhanced metal-poor stars \citep{2008ARA&A..46..241S}. To date, a total of eight $r$-process enhanced metal-poor stars with both thorium (Th, $Z=90$) and uranium (U, $Z=92$) have been detected, including CS~31082-001 \citep{2002A&A...387..560H}, BD~+17{\textdegree}3248 \citep{2002ApJ...572..861C}, HE~1523-0901 \citep{2007ApJ...660L.117F}, CS~29497-004 \citep{2017A&A...607A..91H}, J2038-0023 \citep{2017ApJ...844...18P}, J0954+5246 \citep{2018ApJ...859L..24H}, J2003-1142 \citep{2021Natur.595..223Y}, and J2213-5137 \citep{2024ApJ...971..158R}. According to the Th/U nuclear chronometer, the ages of these stars are consistent with the cosmic age of 13.8 billion years \citep{2022ApJ...941..152W}, suggesting they were polluted by only a single $r$-process event. In this work, we study the detailed heavy element abundances using the $r$-process nuclear reaction network and the astrophysical parameters derived from numerical relativity simulations, and compare these with the observed abundances in $r$-process enhanced metal-poor stars.

The $r$-process enhanced metal-poor stars provide valuable examples for studying the impact of nuclear physics models on $r$-process nucleosynthesis \citep{2009PhRvC..80f5806N}. The path of $r$-process nucleosynthesis nearly reaches the neutron drip line, which is far away from the valley of stability. These extremely neutron-rich nuclei remain beyond current experimental capabilities, and their properties are often unmeasured \citep{2016PrPNP..86...86M,2019JPhG...46h3001H,2019PrPNP.107..109K}. Consequently, theoretical values derived from nuclear physical models are essential for $r$-process calculations. While nuclear physics models generally reproduce experimental values for measured nuclei, predictions for extremely neutron-rich and unmeasured nuclei can vary significantly among different models (see a review by \citealp{2016PrPNP..86...86M} and references therein). As a result, abundance patterns of heavy elements calculated by using different nuclear physics models show significant variation \citep{2015ApJ...808...30E,2015PhRvC..92e5805M,2015PhRvC..92c5807M,2021ApJ...906...94Z,2023MNRAS.520.2806C}, especially in the region of rare earth elements from La ($Z=57$) to Lu ($Z=71$) \citep{2001PhRvC..64c5801S,2011PhRvC..83d5809A,2012PhRvC..86c5803M,2016PrPNP..86...86M,2023PhLB..84438092H}. In the observation of metal-poor stars, heavy elements from Ba ($Z=56$) to Hf ($Z=72$) can be effectively identified from the observed spectrum. Therefore, it is worthwhile to study the impact of nuclear physics models on $r$-process nucleosynthesis and to compare calculated abundances with observed abundances in very metal-poor stars that were polluted by a single $r$-process event. In this work, we select eight $r$-process enhanced metal-poor stars with both Th and U detected, as these actinides are pure $r$-process products that can only be formed via $r$-process nucleosynthesis. We compare the heavy element abundances obtained from different nuclear physics models with those observed in $r$-process enhanced metal-poor stars to explore the impact of nuclear physics inputs on $r$-process nucleosynthesis.

This paper is organized as follows. In Section~\ref{method}, we provide details on the nuclear physics inputs and the $r$-process nucleosynthesis simulations. Section~\ref{result} compares the $r$-process abundances from neutron star mergers with those observed in $r$-process enhanced metal-poor stars. Finally, the summary is provided in Section~\ref{summary}.

\section{Methods}
\label{method}

To study the nucleosynthesis of heavy elements in neutron star mergers, we use the nuclear reaction network code SkyNet to perform $r$-process nucleosynthesis simulations. SkyNet was developed at Los Alamos National Laboratory by \citet{2017ApJS..233...18L} and has been widely used for $r$-process nucleosynthesis in core-collapse supernovae and neutron star mergers. The network contains more than 7,800 nuclide species and includes over 140,000 nuclear reactions. The reaction rates used in SkyNet are taken from the JINA REACLIB database \citep{2010ApJS..189..240C}. Nuclear data are taken from the WebNucleo XML file distributed with REACLIB, and nuclear masses are consistent with the Finite-Range Droplet Model provided by \citet{1995ADNDT..59..185M}.

\begin{figure}[t]
    \centering
    \includegraphics[width=0.45\textwidth]{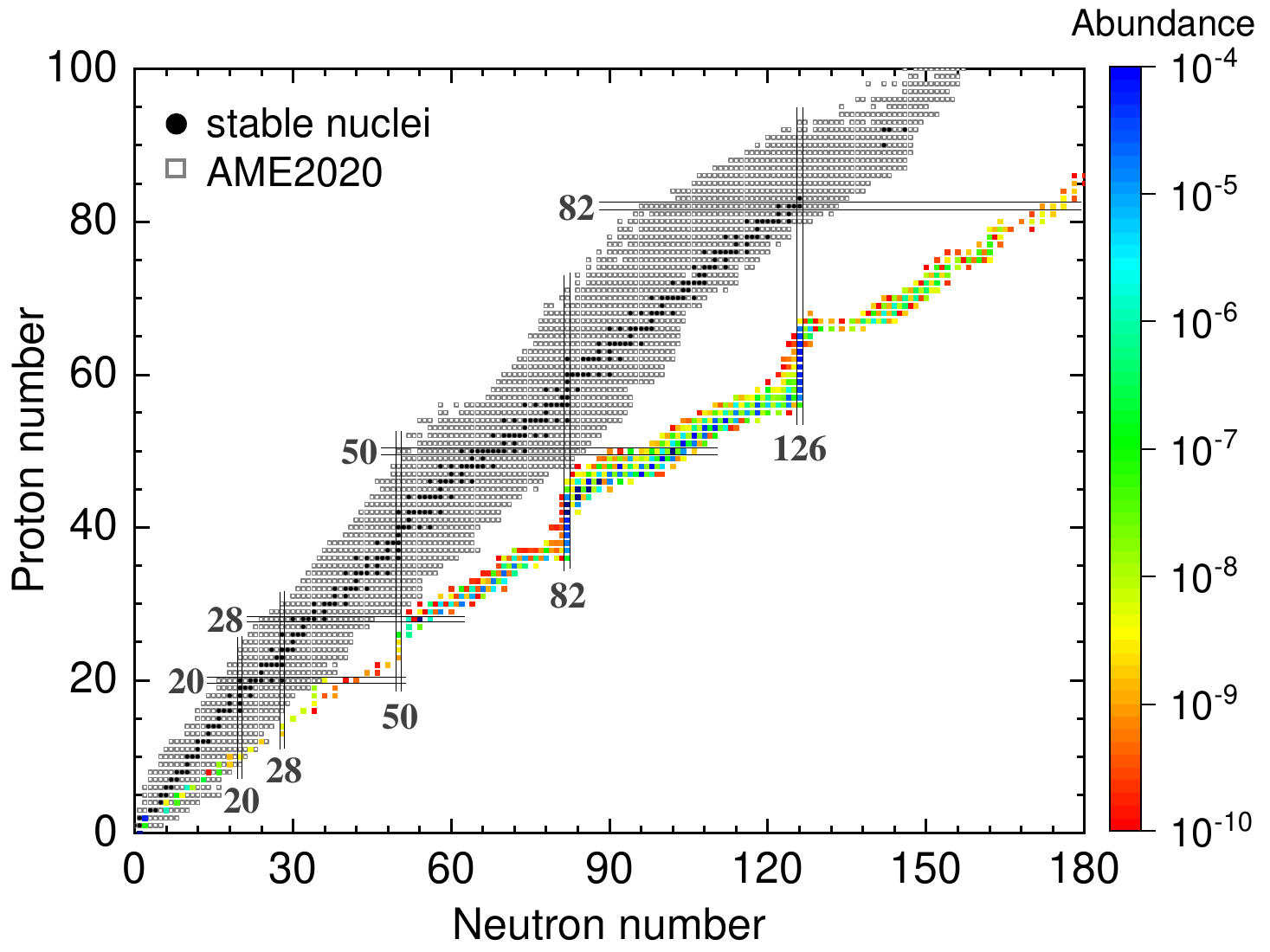}
    \caption{Comparison of typical features of the $r$-process path with the latest measured nuclei from the AME2020 database. The abundance distribution corresponds to the early stages of $r$-process nucleosynthesis at $t=0.2$~s, when the neutron capture process has begun. Our nucleosynthesis calculations start when the temperature drops below $T=6\times10^9$~K. The astrophysical parameters are typical for binary neutron star mergers: electron fraction $Y_{\rm e}=0.10$, specific entropy $s=10$~$k_{\rm B}$/baryon, and expansion timescale $\tau=10$~ms. The black horizontal and vertical lines indicate closed proton and neutron shell nuclei, respectively.}
    \label{path}
\end{figure}

The $r$-process nucleosynthesis simulations depend on theoretical results derived from nuclear physics models. This is caused by the fact that, in the early stages of the $r$-process, a significant amount of extremely neutron-rich nuclei is produced by the neutron capture process. In Figure~\ref{path}, we show the typical features of abundance distribution in the early stages of the $r$-process simulations. For comparison, the latest measured nuclear mass data from AME2020 database \citep{2021ChPhC..45c0003W} are also shown. It can be observed that the $r$-process path lies on the extremely neutron-rich side, far away from the valley of stable nuclei. These extremely neutron-rich nuclei cannot be produced in the laboratory and their properties are unmeasured. Therefore, theoretical predictions from nuclear physics models play a crucial role in $r$-process nucleosynthesis simulations (see recent reviews e.g. by \citealp{2021RvMP...93a5002C,2023A&ARv..31....1A}, and references therein).

\begin{figure*}[t]
    \centering
    \includegraphics[width=0.8\textwidth]{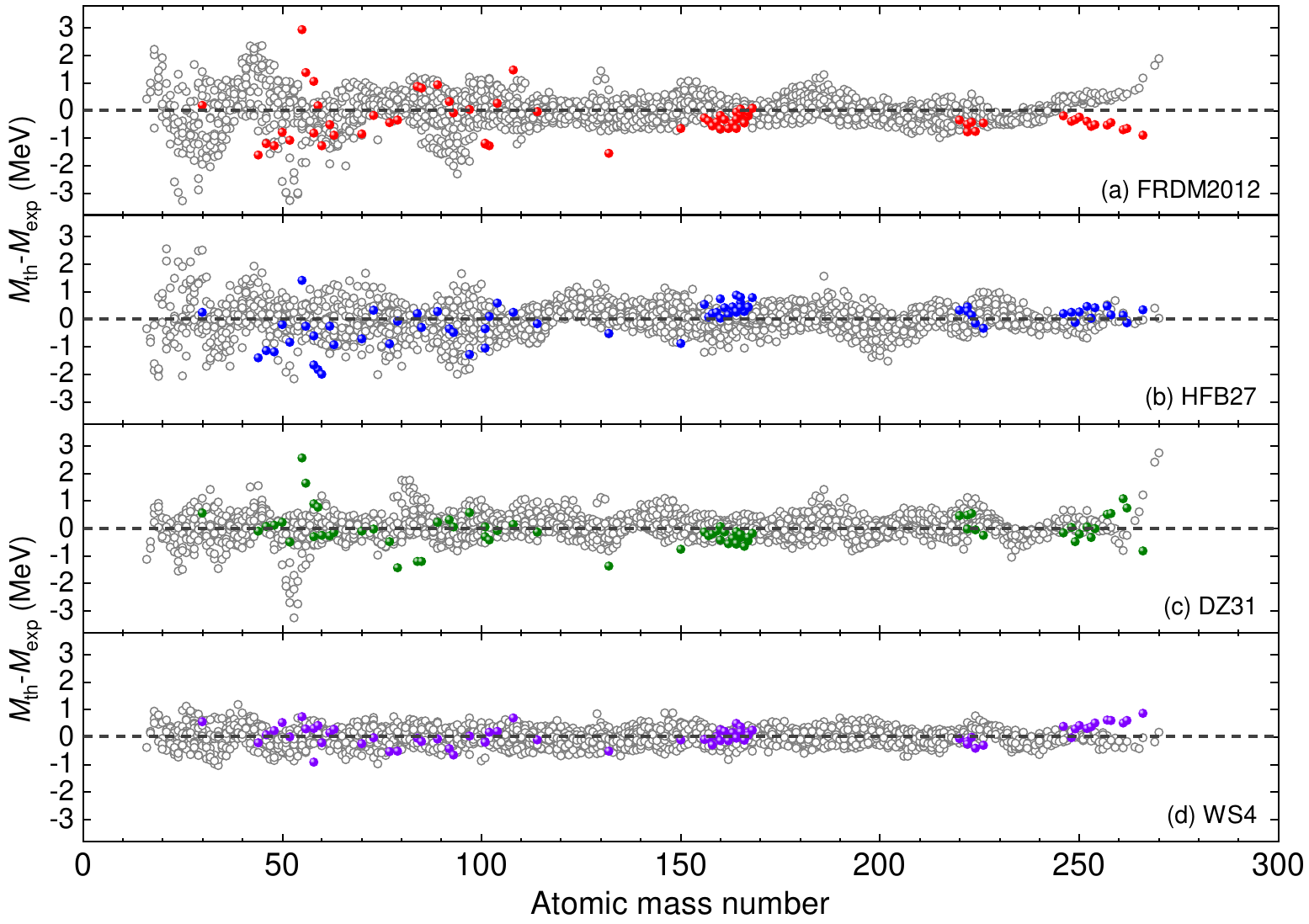}
    \caption{Deviations between theoretical values from nuclear mass models and experimental values in the AME2020 database. Four widely used nuclear mass models are considered: FRDM2012, HFB27, DZ31, and WS4. The solid circles represent newly measured nuclei included in the AME2020 but not in the previous AME2016.}
    \label{mass}
\end{figure*}

In order to reduce the uncertainty caused by nuclear physics inputs, we update the nuclear data with state-of-the-art databases, including the latest nuclear mass data from AME2020 \citep{2021ChPhC..45c0003W} and the latest nuclear decay data from NUBASE2020 \citep{2021ChPhC..45c0001K}. For nuclei without experimental data, we use theoretical values from widely-used nuclear mass models, including Finite-Range Droplet Model (FRDM2012; \citealp{2012PhRvL.108e2501M,2016ADNDT.109....1M}), Hartree-Fock-Bogoliubov (HFB27; \citealp{2009PhRvL.102o2503G,2013PhRvC..88f1302G}), Duflo-Zuker (DZ31; \citealp{1995PhRvC..52...23D}), and Weizs$\ddot{\rm a}$cker-Skyrme (WS4; \citealp{2010PhRvC..82d4304W,2014PhLB..734..215W}). The FRDM2012 is the most popular model based on the macroscopic-microscopic approach, where the macroscopic part treats the nucleus as a deformed liquid droplet, and the microscopic part accounts for nuclear shell correction and pairing effect \citep{2016ADNDT.109....1M}. The HFB27 model, which is a leading example of a fully microscopic approach, predicts nuclear masses based on the Skyrme energy density functional theory \citep{2013PhRvC..88f1302G}. The DZ31 model is an empirical formula that successfully describes nuclear masses across a wide range of nuclei \citep{1995PhRvC..52...23D}. The WS4 model is the most recent and improved version of the macroscopic-microscopic approach, which uses the Skyrme energy density functional and accounts for the surface diffuseness effect in neutron-rich nuclei \citep{2014PhLB..734..215W}. These four nuclear physics models can predict nuclear masses for extremely neutron-rich nuclei and have been widely used in $r$-process simulations.

Figure~\ref{mass} presents the deviations between theoretical nuclear mass values predicted by various models and experimental values reported in the AME2020 database. Solid circles indicate the newly measured nuclei included in the AME2020 database but absent from the previous AME2016 database \citep{2017ChPhC..41c0003W}. These newly measured data provide a unique opportunity to test the predictive power of nuclear physics models, as theoretical models had predicted these masses prior to measurement. The deviations between theoretical mass models and experimental data are generally within 3~MeV. The nuclear masses predicted by the WS4 model exhibit better agreement with the experimental data compared to those from the other three mass models. This indicates that the WS4 model is both powerful and effective in predicting nuclear masses for unknown nuclei.

To quantify the predictive power of nuclear physics models, we calculate the root-mean-square (rms) deviation between theoretical and experimental values as
\begin{equation}
    \sigma(M) = \sqrt{ \frac{1}{n} \sum_{1}^{n} \left(M^{\rm th}_i - M^{\rm exp}_i\right)^2 },
\end{equation}
where $M^{\rm th}_i$ and $M^{\rm exp}_i$ are the theoretical and experimental values of the $i$th nuclear species, respectively, and $n$ is the total number of nuclide species. Table~\ref{RMS} lists the rms deviations between theoretical mass values from various models and experimental values from the AME2020 database.

\begin{table}[h]
    \centering
    \caption{Comparison of rms deviations between theoretical values from nuclear physics models and experimental data in the AME2020 database.}
    \begin{tabular}{c|cc|cc}
    \hline\hline
    \multirow{2}{*}{Model}  &  $\sigma(M)_{\rm tot}$  &  $\sigma(M)_{\rm new}$  &  \multirow{2}{*}{$\sigma(\lambda)_{\rm tot}$}  &  \multirow{2}{*}{$\sigma(\lambda)_{\rm new}$} \\
       &   [MeV]   &   [MeV]   &   &   \\
    \hline
    FRDM2012    &   0.606   &   0.796   &   0.385   &   0.426   \\
    HFB27   &   0.518   &   0.673   &   0.506   &   0.323   \\
    DZ31    &   0.425   &   0.619   &   0.376   &   0.281   \\
    WS4     &   0.295   &   0.360   &   0.318   &   0.331   \\
    \hline
    \end{tabular}
    \label{RMS}
    \tablefoot{The second and third columns represent deviations in nuclear mass, while the fourth and fifth columns represent deviations in neutron capture rate. Columns labeled ``tot'' include all nuclide species present in the AME2020 database, whereas columns labeled ``new'' include newly measured species present in AME2020 but not in AME2016.}
\end{table}

\begin{figure*}[t]
    \centering
    \includegraphics[width=0.8\textwidth]{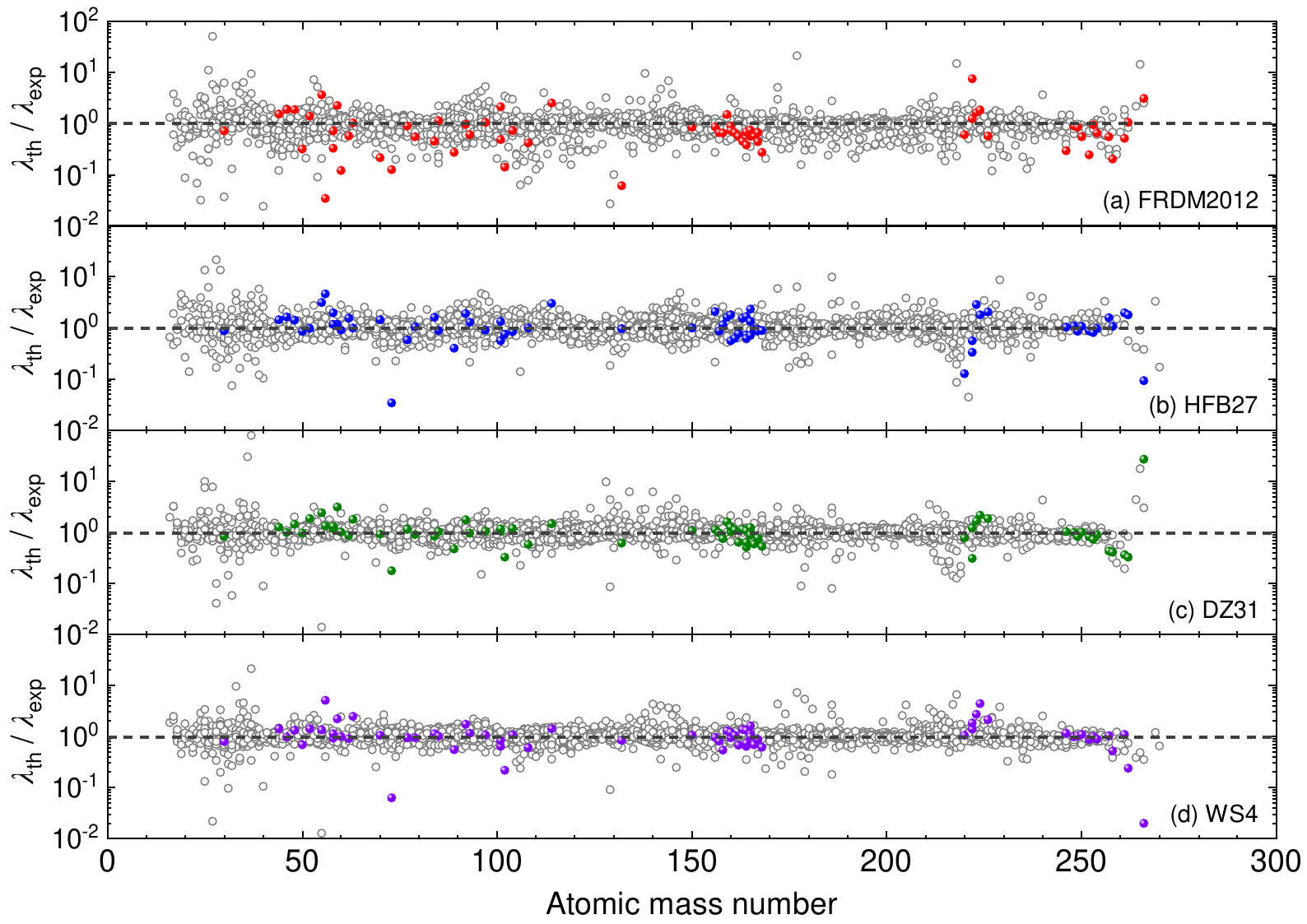}
    \caption{The same as Figure~\ref{mass}, but showing deviations for neutron capture rates at a temperature of $10^{9}$~K.}
    \label{capture1}
\end{figure*}

The reaction rates of neutron capture process (n, $\gamma$) are recalculated using updated nuclear masses with the Hauser-Feshbach statistical code TALYS \citep{2008A&A...487..767G}. TALYS provides a detailed description of astrophysical reaction rates in the temperature range from $10^{8}$ to $10^{10}$~K. The calculated reaction rates $\lambda$ are then converted to a polynomial format consistent with the REACLIB database
\begin{equation}
    \lambda = \exp \left({a_0 + a_1 T_9^{-1} + a_2 T_9^{-1/3} + a_3 T_9^{1/3} + a_4 T_9 + a_5 T_9^{5/3} + a_6 \ln{T_9} }\right),
\end{equation}
where $T_9 = 10^9$~K, and the coefficients $a_0 \sim a_6$ are obtained using the least squares fitting method in logarithmic space. The inverse reaction rates ($\gamma$, n) are calculated by detailed balance. Nuclear fission rates are calculated using fission barriers from \cite{2015PhRvC..91b4310M} and fission fragment distributions from \cite{1975NuPhA.239..489K}.

The rms deviation between neutron capture rates based on theoretical models and those based on experimental data can be written as
\begin{equation}
    \sigma(\lambda) = \sqrt{ \frac{1}{n} \sum_{1}^{n} \left(\log{\lambda^{\rm th}_i} - \log{\lambda^{\rm exp}_i} \right)^2 },
\end{equation}
where $\lambda^{\rm th}_i$ and $\lambda^{\rm exp}_i$ are the theoretical and experimental neutron capture rates for the $i$th nuclide species, respectively.

Since this study focuses on the impact of nuclear physics inputs on $r$-process nucleosynthesis simulations, we adopt parameterized trajectories as astrophysical inputs. Astrophysical parameters are taken from numerical relativity simulations provided by \citet{2018ApJ...869..130R}. Table~\ref{input} lists the astrophysical parameters adopted in our $r$-process nucleosynthesis simulations. The astrophysical parameters for disk wind ejecta are broadly consistent with those in \cite{2021ApJ...906...98N}, with an electron fraction of $Y_{\rm e} = 0.3$, a specific entropy of $s=20~k_{\rm B}$/baryon, and an expansion velocity of $v=0.15c$. The density profile adopts the analytical expression consistent with \cite{2015ApJ...815...82L}, which initially decreases exponentially with time, i.e., $\rho \propto e^{-t}$, and then smoothly transitions to a homologous expansion, $\rho \propto t^{-3}$. The evolution of temperature is influenced by both the nuclear equation of state and self-heating from nuclear reactions. In our simulations, the $r$-process starts when the temperature drops below $6\times10^9$~K and ends at $t=10^9$~s, by which time the majority of radioactive nuclei have decayed into stable elements.

\begin{table}
    \centering
    \caption{Astrophysical parameters for $r$-process simulations.}
    \begin{tabular}{cccccc}
    \hline\hline
    \multirow{2}{*}{Label}  &  \multirow{2}{*}{$Y_{\rm e}$}  &  $s$  &  $v$  \\
       &   &   [$k_{\rm B}$]   &   [c]   \\
    \hline
    BHBlp\_M135135    &   0.15    &   20  &   0.17   \\
    BHBlp\_M140140    &   0.15    &   18  &   0.17   \\
    DD2\_M130130      &   0.13    &   15  &   0.18   \\
    DD2\_M140140      &   0.17    &   22  &   0.22   \\
    LS220\_M140140    &   0.14    &   16  &   0.17   \\
    LS220\_M144139    &   0.14    &   15  &   0.16   \\
    \hline
    \end{tabular}
    \tablefoot{Astrophysical parameters are taken from numerical relativity simulations provided by \citet{2018ApJ...869..130R}, including the electron fraction $Y_{\rm e}$, specific entropy $s$, and expansion velocity $v$.}
    \label{input}
\end{table}

The rms deviation for the abundances of $r$-process elements is given by
\begin{equation}
    \sigma(Y) = \sqrt{ \frac{1}{n} \sum_{1}^{n} \left(\log{Y_i^{\rm cal}} - \log{Y_i^{\rm obs}} \right)^2 },
\end{equation}
where $Y_i^{\rm cal}$ is the calculated abundances using nuclear physics models, and $Y_i^{\rm obs}$ is the observed abundances in $r$-process enhanced metal-poor stars.

\section{Results and analysis}
\label{result}

Figure~\ref{capture1} shows the deviations of neutron capture rates calculated using theoretical mass values from various models compared to those based on experimental values from the AME2020 database. The neutron capture rates calculated with the WS4 model are in good agreement with those based on experimental values, and the WS4 model's results are significantly more accurate than those from the other three nuclear physics models. The deviations of neutron capture rates with the WS4 model are within the same order of magnitude as the experimental values, including those for newly measured nuclei.

\begin{figure}
    \centering
    \includegraphics[width=0.45\textwidth]{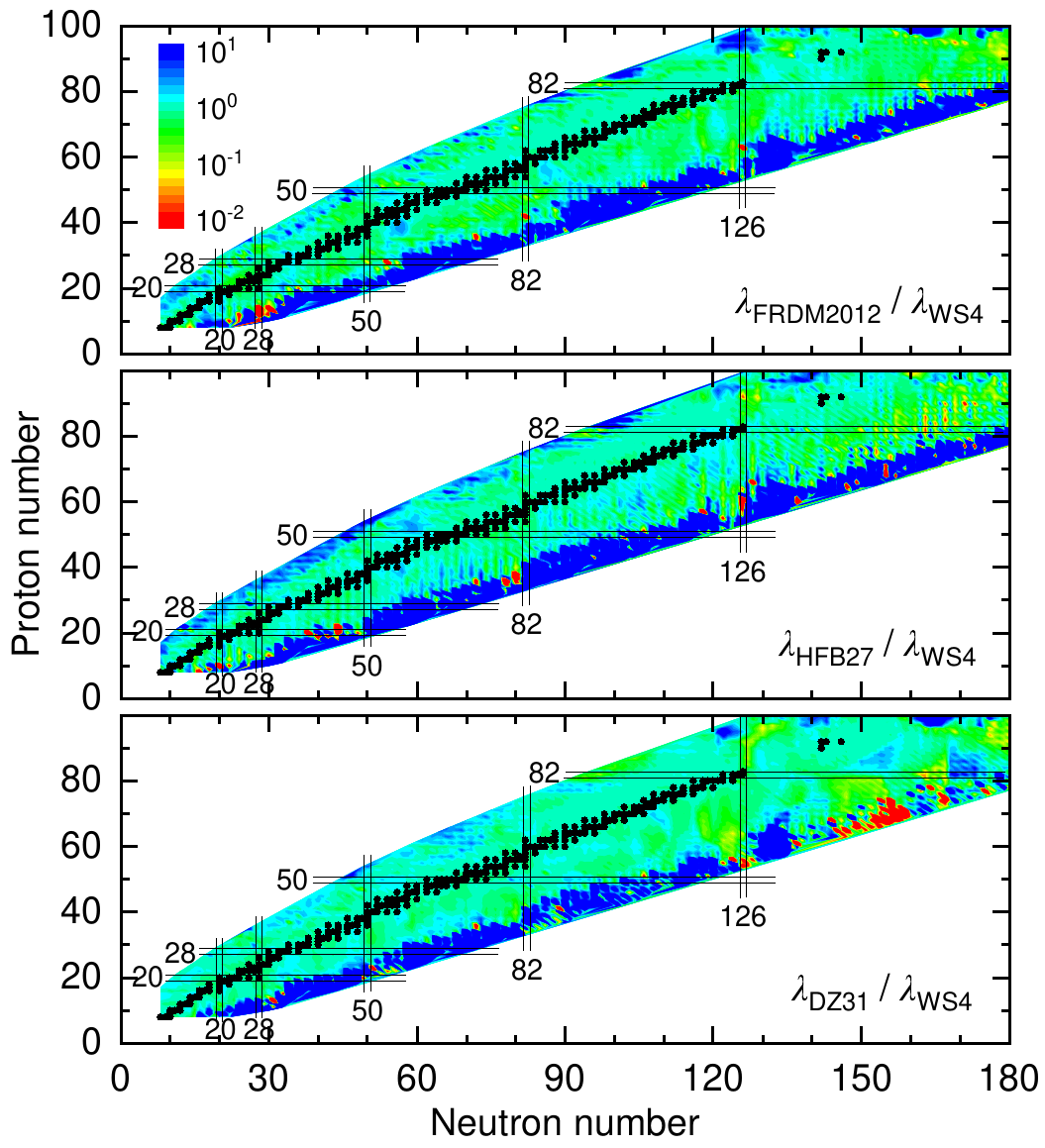}
    \caption{Comparison of neutron capture rates for each nucleus calculated using various nuclear physics models.}
    \label{capture2}
\end{figure}

To further analyze the neutron capture rates calculated using various nuclear physics models, we compare the results for each nuclide species, as shown in Figure~\ref{capture2}. In the region of measured nuclei, neutron capture rates from different models are generally consistent. However, in the region approaching the $r$-process path, neutron capture rates calculated using different models exhibit significant deviations, reaching up to one order of magnitude. Table~\ref{RMS} lists the rms deviations of neutron capture rates based on different nuclear physics models compared to those based on experimental values. It is found that the rms deviations of the WS4 model for both total and newly measured nuclei are approximately 0.3, indicating that the overall discrepancy between neutron capture rates based on theoretical values and those based on experimental data is within a factor of $10^{0.3} \approx 2$. 

\begin{figure}
    \centering
    \includegraphics[width=0.5\textwidth]{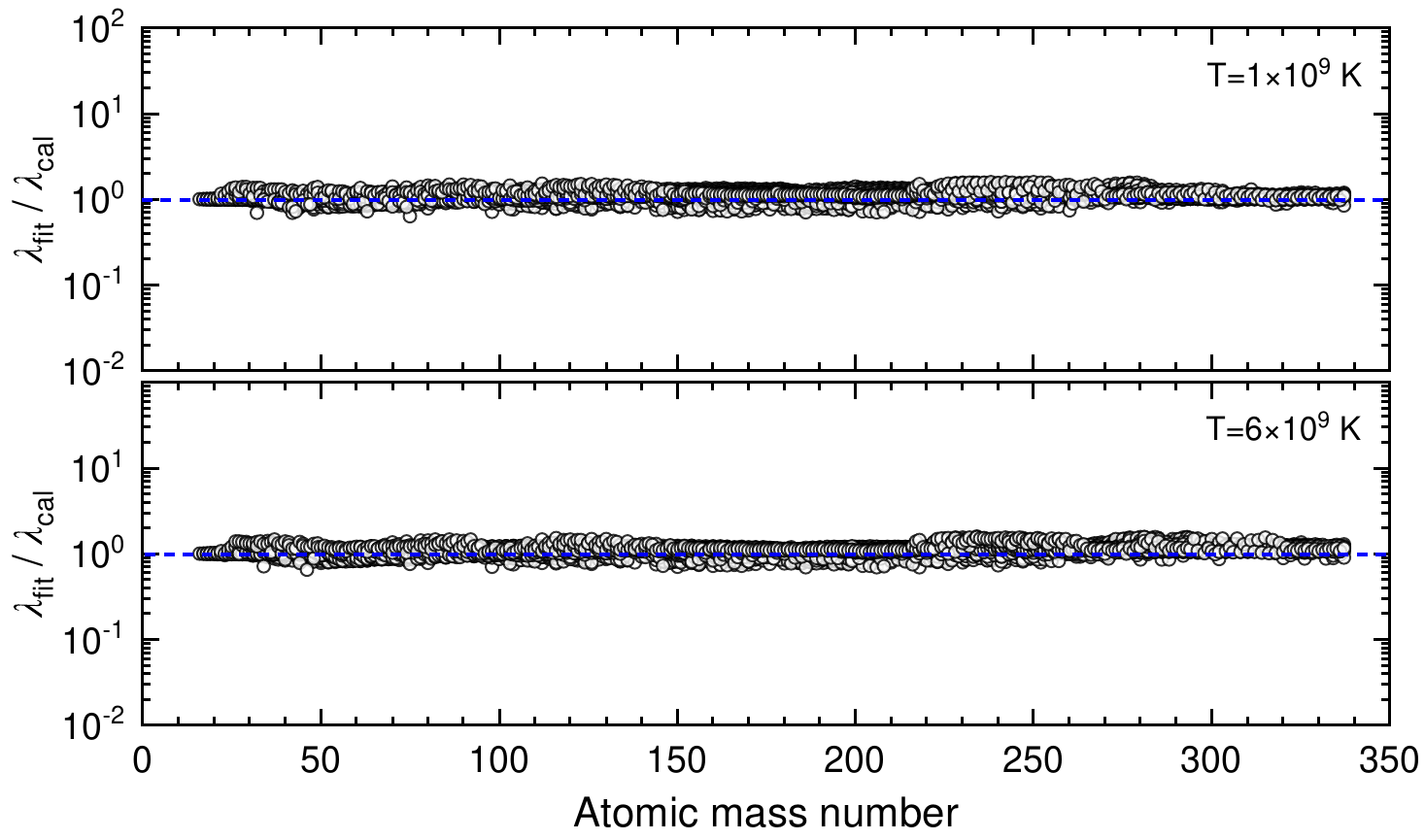}
    \caption{Deviations between the neutron capture rates obtained using the fitted polynomial format and those calculated with the TALYS code. The neutron capture rates are shown at temperatures of $1\times10^9$~K (top panel) and $6\times10^9$~K (bottom panel). Nuclear mass values are based on the WS4 model.}
    \label{fitting}
\end{figure}

Since we have converted the neutron capture rates obtained using the TALYS method into a polynomial format consistent with the REACLIB database, it is necessary to examine the deviations between the fitting results and the original TALYS calculations. Figure~\ref{fitting} shows the deviations between the neutron capture rates obtained from the fitted polynomial and those directly calculated using the TALYS code. The neutron capture rates shown in the figure are based on nuclear mass values from the WS4 model. It can be observed that the neutron capture rates derived from the fitted polynomial are in good agreement with those directly calculated by the TALYS code.

\begin{figure*}
    \centering
    \includegraphics[width=0.8\textwidth]{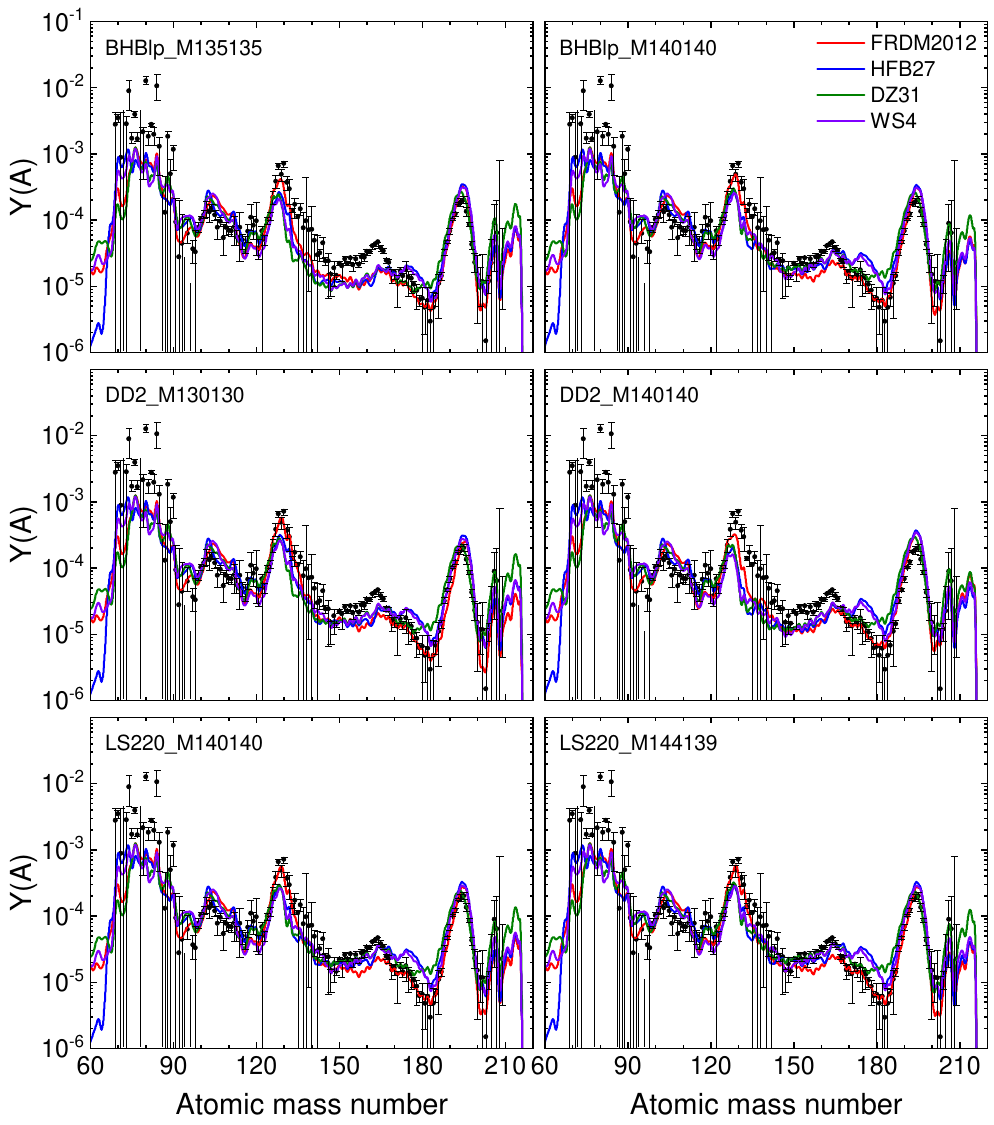}
    \caption{Abundance patterns of $r$-process elements produced by neutron star mergers at $t=10^9$~s. The solar $r$-process abundances from \cite{2007PhR...450...97A} are shown for comparison. Astrophysical parameters used in the simulations are detailed in Table~\ref{input}.}
    \label{YA}
\end{figure*}

We use updated nuclear physics inputs to explore their impact on $r$-process nucleosynthesis simulations. Figure~\ref{YA} shows the abundance patterns of $r$-process elements as simulated with different nuclear physics inputs. For comparison, the solar $r$-process abundances from  \cite{2007PhR...450...97A} are shown. Astrophysical parameters for $r$-process simulations are listed in Table~\ref{input}. It can be seen that the abundance patterns from $r$-process simulations are broadly consistent with the solar $r$-process abundances, particularly in the regions of the second and third $r$-process peaks. However, for nuclei in the region of rare earth elements (with atomic mass numbers $140\leq A \leq 180$), the abundances calculated using the FRDM2012 model are lower than those from the other three nuclear physics models. This discrepancy may be attributed to the significant differences in nuclear mass values and neutron capture rates obtained using the FRDM2012 model compared to those calculated with the other three nuclear physics models. To determine the impact of individual nuclear properties on $r$-process simulations, it is necessary to vary each nuclear property and study its sensitivity to element abundances \citep{2016PrPNP..86...86M}.

\begin{figure*}
    \centering
    \includegraphics[width=0.9\textwidth]{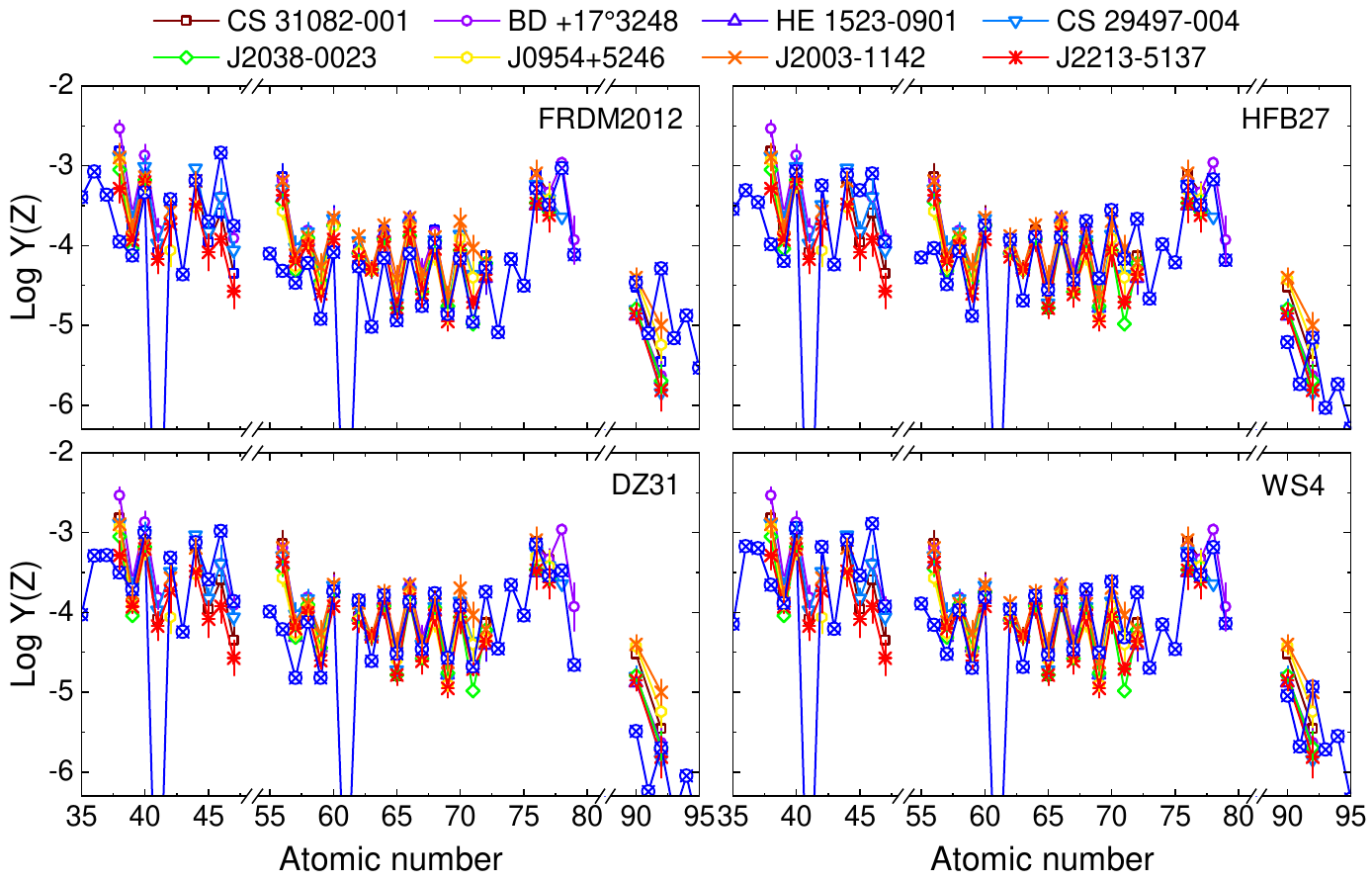}
    \caption{Comparison of $r$-process abundances from neutron star mergers with those observed in $r$-process enhanced metal-poor stars. Eight $r$-process enhanced metal-poor stars are considered: CS~31082-001 \citep{2002A&A...387..560H}, BD~+17{\textdegree}3248 \citep{2002ApJ...572..861C}, HE~1523-0901 \citep{2007ApJ...660L.117F}, CS~29497-004 \citep{2017A&A...607A..91H}, J2038-0023 \citep{2017ApJ...844...18P}, J0954+5246 \citep{2018ApJ...859L..24H}, J2003-1142 \citep{2021Natur.595..223Y}, and J2213-5137 \citep{2024ApJ...971..158R}. The element abundances of these stars are scaled to $Y(Z=63)=5.0\times10^{-5}$. The blue lines represent the average abundances across the astrophysical models for neutron star mergers.}
    \label{MPS}
\end{figure*}

The $r$-process enhanced metal-poor stars provide robust cases for studying the impact of nuclear physics models on $r$-process simulations, particularly for nuclei in the rare earth element region. We compare the $r$-process abundances from neutron star mergers with those observed in $r$-process enhanced metal-poor stars, as shown in Figure~\ref{MPS}. We note that it is difficult to distinguish between different isotopes in the observations of metal-poor stars, and thus the x-axis in the figure represents atomic number rather than atomic mass number. The element abundances of $r$-process enhanced metal-poor stars are scaled to $Y(Z=63)=5.0\times10^{-5}$. It can be seen that the $r$-process abundances calculated using the FRDM2012 model are generally lower than those observed in $r$-process enhanced metal-poor stars, except for Th and U. The observed abundances of metal-poor stars are well reproduced by the $r$-process simulations using the WS4 model. Additionally, the odd-even effect in the heavy element abundances of $r$-process enhanced metal-poor stars is also well reproduced, where the abundances of odd-Z nuclei are slightly lower than those of their neighboring even-Z nuclei. In Table~\ref{RMS_MPS}, we present the rms deviations between the simulated abundances from various nuclear physics models and the observed abundances in $r$-process enhanced metal-poor stars. It is found that using the WS4 model in $r$-process nucleosynthesis simulations can more accurately reproduce the $r$-process abundances observed in metal-poor stars compared to other three nuclear physics models.

\begin{table}
    \centering
    \caption{Comparison of rms deviations between simulated abundances from various nuclear physics models and observed abundances in $r$-process enhanced metal-poor stars.}
    \begin{tabular}{cc}
    \hline\hline
    Model   &   $\sigma(Y)$  \\
    \hline
    FRDM2012    &   0.456   \\
    HFB27       &   0.380   \\
    DZ31        &   0.372   \\
    WS4         &   0.347   \\
    \hline
    \end{tabular}
    \label{RMS_MPS}
    \tablefoot{Here we compare the average abundances from six astrophysical models for neutron star mergers with the average abundances observed in eight $r$-process enhanced metal-poor stars.}
\end{table}

\section{Summary}
\label{summary}

The rapid neutron-capture process ($r$-process) involves a significant number of extremely neutron-rich and unmeasured nuclei, and thus relies heavily on theoretical predictions provided by nuclear physics models (Figure~\ref{path}). Several metal-poor stars have been observed with detailed composition of $r$-process elements, offering valuable cases for studying the impact of nuclear physics inputs on $r$-process nucleosynthesis. In this work, we conducted a detailed comparison between the $r$-process abundances calculated using a nuclear reaction network with various nuclear physics models and the observed abundances in $r$-process enhanced metal-poor stars, providing insights into the predictive power of these models and their impact on $r$-process nucleosynthesis.

Firstly, we compared theoretical values from nuclear physics models with experimental values from the latest AME2020 database. It was found that nuclear masses predicted by the Weizs$\ddot{\rm a}$cker-Skyrme (WS4) model are in good agreement with experimental data, especially for the newly measured nuclei (Figure~\ref{mass}). The root-mean-square (rms) deviation between theoretical values from the WS4 model and experimental values from the AME2020 database is $\sim0.3$~MeV, which is significantly smaller than those from other nuclear mass models. This indicates that the WS4 model is both powerful and effective in predicting nuclear masses for unknown nuclei.

Secondly, we analyzed the neutron capture rates calculated using theoretical values from various nuclear physics models and compared them with those based on experimental data. In the region of measured nuclei, neutron capture rates from different models are generally consistent. However, in the region approaching the $r$-process path, neutron capture rates calculated using different models exhibit significant deviations, reaching up to one order of magnitude (Figure~\ref{capture2}). Neutron capture rates calculated with the WS4 model are in good agreement with those based on experimental values (Figure~\ref{capture1}).

Finally, we studied the $r$-process abundances using a nuclear reaction network with updated nuclear physics inputs and compared these results with the abundances observed in $r$-process enhanced metal-poor stars. It was found that the WS4 model provides a more accurate reproduction of $r$-process abundances observed in metal-poor stars compared to other three nuclear physics models (Figure~\ref{MPS}), particularly in the region of rare earth elements from La ($Z=57$) to Lu ($Z=71$). The odd-even effect in the heavy element abundances of $r$-process enhanced metal-poor stars is also well reproduced, where the abundances of odd-Z nuclei are slightly lower than those of their neighboring even-Z nuclei. This may be due to the fact that the nuclear physics inputs (i.e., nuclear mass values and neutron capture rates) derived from the WS4 model are in good agreement with those based on experimental data. These results suggest that the WS4 model plays a crucial role in $r$-process nucleosynthesis simulations, especially in studies that require quantitative analysis of elemental abundances, such as age determinations of metal-poor stars using Th/U nuclear chronometry \citep{2022ApJ...941..152W}.

In summary, the WS4 model not only provides precise predictions for nuclear properties but also plays a crucial role in $r$-process nucleosynthesis simulations. Therefore, we recommend using the theoretical values from the WS4 model as nuclear physics inputs for $r$-process nucleosynthesis simulations. The nuclear mass table for the WS4 model, provided by \cite{2014PhLB..734..215W}, is available at \href{http://www.imqmd.com/mass/}{http://www.imqmd.com/mass/}.

\begin{acknowledgements}

We thank Hui-Ling Chen for valuable discussions.
This work was supported by the National Natural Science Foundation of China (Grant Nos. 11973014, 12133003, 12347172, 12403043, and 12473038). M.H.C. also acknowledges support from the China Postdoctoral Science Foundation (Grant Nos. GZB20230029 and 2024M750057). This work was also supported by the Guangxi Talent Program (Highland of Innovation Talents).

\end{acknowledgements}

\begin{center}
ORCID iDs
\end{center}

Meng-Hua Chen: https://orcid.org/0000-0001-8406-8683

Li-Xin Li: https://orcid.org/0000-0002-8466-321X

En-Wei Liang: https://orcid.org/0000-0002-7044-733X

Ning Wang: https://orcid.org/0000-0002-2629-6730


\begin{thebibliography}{64}
\expandafter\ifx\csname natexlab\endcsname\relax\def\natexlab#1{#1}\fi

\bibitem[{{Abbott} {et~al.}(2017){Abbott}, {Abbott}, {Abbott}, {Acernese}, {Ackley}, {Adams}, {Adams}, {Addesso}, {Adhikari}, {Adya}, {Affeldt}, {Afrough}, {Agarwal}, {Agathos}, {Agatsuma}, {Aggarwal}, {Aguiar}, {Aiello}, {Ain}, {Ajith}, {Allen}, {Allen}, {Allocca}, {Altin}, {Amato}, {Ananyeva}, {Anderson}, {Anderson}, {Angelova}, {Antier}, {Appert}, {Arai}, {Araya}, {Areeda}, {Arnaud}, {Arun}, {Ascenzi}, {Ashton}, {Ast}, {Aston}, {Astone}, {Atallah}, {Aufmuth}, {Aulbert}, {AultONeal}, {Austin}, {Avila-Alvarez}, {Babak}, {Bacon}, {Bader}, {Bae}, {Baker}, {Baldaccini}, {Ballardin}, {Ballmer}, {Banagiri}, {Barayoga}, {Barclay}, {Barish}, {Barker}, {Barkett}, {Barone}, {Barr}, {Barsotti}, {Barsuglia}, {Barta}, {Barthelmy}, {Bartlett}, {Bartos}, {Bassiri}, {Basti}, {Batch}, {Bawaj}, {Bayley}, {Bazzan}, {B{\'e}csy}, {Beer}, {Bejger}, {Belahcene}, {Bell}, {Berger}, {Bergmann}, {Bero}, {Berry}, {Bersanetti}, {Bertolini}, {Betzwieser}, {Bhagwat}, {Bhandare}, {Bilenko}, {Billingsley}, {Billman}, {Birch}, {Birney},
  {Birnholtz}, {Biscans}, {Biscoveanu}, {Bisht}, {Bitossi}, {Biwer}, {Bizouard}, {Blackburn}, {Blackman}, {Blair}, {Blair}, {Blair}, {Bloemen}, {Bock}, {Bode}, {Boer}, {Bogaert}, {Bohe}, {Bondu}, {Bonilla}, {Bonnand}, {Boom}, {Bork}, {Boschi}, {Bose}, {Bossie}, {Bouffanais}, {Bozzi}, {Bradaschia}, {Brady}, {Branchesi}, {Brau}, {Briant}, {Brillet}, {Brinkmann}, {Brisson}, {Brockill}, {Broida}, {Brooks}, {Brown}, {Brown}, {Brunett}, {Buchanan}, {Buikema}, {Bulik}, {Bulten}, {Buonanno}, {Buskulic}, {Buy}, {Byer}, {Cabero}, {Cadonati}, {Cagnoli}, {Cahillane}, {Calder{\'o}n Bustillo}, {Callister}, {Calloni}, {Camp}, {Canepa}, {Canizares}, {Cannon}, {Cao}, {Cao}, {Capano}, {Capocasa}, {Carbognani}, {Caride}, {Carney}, {Casanueva Diaz}, {Casentini}, {Caudill}, {Cavagli{\`a}}, {Cavalier}, {Cavalieri}, {Cella}, {Cepeda}, {Cerd{\'a}-Dur{\'a}n}, {Cerretani}, {Cesarini}, {Chamberlin}, {Chan}, {Chao}, {Charlton}, {Chase}, {Chassande-Mottin}, {Chatterjee}, {Chatziioannou}, {Cheeseboro}, {Chen}, {Chen}, {Chen}, {Cheng},
  {Chia}, {Chincarini}, {Chiummo}, {Chmiel}, {Cho}, {Cho}, {Chow}, {Christensen}, {Chu}, {Chua}, {Chua}, {Chung}, {Chung}, {Ciani}, {Ciolfi}, {Cirelli}, {Cirone}, {Clara}, {Clark}, {Clearwater}, {Cleva}, {Cocchieri}, {Coccia}, {Cohadon}, {Cohen}, {Colla}, {Collette}, {Cominsky}, {Constancio}, {Conti}, {Cooper}, {Corban}, {Corbitt}, {Cordero-Carri{\'o}n}, {Corley}, {Cornish}, {Corsi}, {Cortese}, {Costa}, {Coughlin}, {Coughlin}, {Coulon}, {Countryman}, {Couvares}, {Covas}, {Cowan}, {Coward}, {Cowart}, {Coyne}, {Coyne}, {Creighton}, {Creighton}, {Cripe}, {Crowder}, {Cullen}, {Cumming}, {Cunningham}, {Cuoco}, {Dal Canton}, {D{\'a}lya}, {Danilishin}, {D'Antonio}, {Danzmann}, {Dasgupta}, {Da Silva Costa}, {Dattilo}, {Dave}, {Davier}, {Davis}, {Daw}, {Day}, {De}, {DeBra}, {Degallaix}, {De Laurentis}, {Del{\'e}glise}, {Del Pozzo}, {Demos}, {Denker}, {Dent}, {De Pietri}, {Dergachev}, {De Rosa}, {DeRosa}, {De Rossi}, {DeSalvo}, {de Varona}, {Devenson}, {Dhurandhar}, {D{\'\i}az}, {Di Fiore}, {Di Giovanni}, {Di
  Girolamo}, {Di Lieto}, {Di Pace}, {Di Palma}, {Di Renzo}, {Doctor}, {Dolique}, {Donovan}, {Dooley}, {Doravari}, {Dorrington}, {Douglas}, {Dovale {\'A}lvarez}, {Downes}, {Drago}, {Dreissigacker}, {Driggers}, {Du}, {Ducrot}, {Dupej}, {Dwyer}, {Edo}, {Edwards}, {Effler}, {Ehrens}, {Eichholz}, {Eikenberry}, {Eisenstein}, {Essick}, {Estevez}, {Etienne}, {Etzel}, {Evans}, {Evans}, {Factourovich}, {Fafone}, {Fair}, {Fairhurst}, {Fan}, {Farinon}, {Farr}, {Farr}, {Fauchon-Jones}, {Favata}, {Fays}, {Fee}, {Fehrmann}, {Feicht}, {Fejer}, {Fernandez-Galiana}, {Ferrante}, {Ferreira}, {Ferrini}, {Fidecaro}, {Finstad}, {Fiori}, {Fiorucci}, {Fishbach}, {Fisher}, {Fitz-Axen}, {Flaminio}, {Fletcher}, {Fong}, {Font}, {Forsyth}, {Forsyth}, {Fournier}, {Frasca}, {Frasconi}, {Frei}, {Freise}, {Frey}, {Frey}, {Fries}, {Fritschel}, {Frolov}, {Fulda}, {Fyffe}, {Gabbard}, {Gadre}, {Gaebel}, {Gair}, {Gammaitoni}, {Ganija}, {Gaonkar}, {Garcia-Quiros}, {Garufi}, {Gateley}, {Gaudio}, {Gaur}, {Gayathri}, {Gehrels}, {Gemme}, {Genin},
  {Gennai}, {George}, {George}, {Gergely}, {Germain}, {Ghonge}, {Ghosh}, {Ghosh}, {Ghosh}, {Giaime}, {Giardina}, {Giazotto}, {Gill}, {Glover}, {Goetz}, {Goetz}, {Gomes}, {Goncharov}, {Gonz{\'a}lez}, {Gonzalez Castro}, {Gopakumar}, {Gorodetsky}, {Gossan}, {Gosselin}, {Gouaty}, {Grado}, {Graef}, {Granata}, {Grant}, {Gras}, {Gray}, {Greco}, {Green}, {Gretarsson}, {Griswold}, {Groot}, {Grote}, {Grunewald}, {Gruning}, {Guidi}, {Guo}, {Gupta}, {Gupta}, {Gushwa}, {Gustafson}, {Gustafson}, {Halim}, {Hall}, {Hall}, {Hamilton}, {Hammond}, {Haney}, {Hanke}, {Hanks}, {Hanna}, {Hannam}, {Hannuksela}, {Hanson}, {Hardwick}, {Harms}, {Harry}, {Harry}, {Hart}, {Haster}, {Haughian}, {Healy}, {Heidmann}, {Heintze}, {Heitmann}, {Hello}, {Hemming}, {Hendry}, {Heng}, {Hennig}, {Heptonstall}, {Heurs}, {Hild}, {Hinderer}, {Hoak}, {Hofman}, {Holt}, {Holz}, {Hopkins}, {Horst}, {Hough}, {Houston}, {Howell}, {Hreibi}, {Hu}, {Huerta}, {Huet}, {Hughey}, {Husa}, {Huttner}, {Huynh-Dinh}, {Indik}, {Inta}, {Intini}, {Isa}, {Isac}, {Isi},
  {Iyer}, {Izumi}, {Jacqmin}, {Jani}, {Jaranowski}, {Jawahar}, {Jim{\'e}nez-Forteza}, {Johnson}, {Jones}, {Jones}, {Jonker}, {Ju}, {Junker}, {Kalaghatgi}, {Kalogera}, {Kamai}, {Kandhasamy}, {Kang}, {Kanner}, {Kapadia}, {Karki}, {Karvinen}, {Kasprzack}, {Katolik}, {Katsavounidis}, {Katzman}, {Kaufer}, {Kawabe}, {K{\'e}f{\'e}lian}, {Keitel}, {Kemball}, {Kennedy}, {Kent}, {Key}, {Khalili}, {Khan}, {Khan}, {Khan}, {Khazanov}, {Kijbunchoo}, {Kim}, {Kim}, {Kim}, {Kim}, {Kim}, {Kim}, {Kimbrell}, {King}, {King}, {Kinley-Hanlon}, {Kirchhoff}, {Kissel}, {Kleybolte}, {Klimenko}, {Knowles}, {Koch}, {Koehlenbeck}, {Koley}, {Kondrashov}, {Kontos}, {Korobko}, {Korth}, {Kowalska}, {Kozak}, {Kr{\"a}mer}, {Kringel}, {Krishnan}, {Kr{\'o}lak}, {Kuehn}, {Kumar}, {Kumar}, {Kumar}, {Kuo}, {Kutynia}, {Kwang}, {Lackey}, {Lai}, {Landry}, {Lang}, {Lange}, {Lantz}, {Lanza}, {Larson}, {Lartaux-Vollard}, {Lasky}, {Laxen}, {Lazzarini}, {Lazzaro}, {Leaci}, {Leavey}, {Lee}, {Lee}, {Lee}, {Lee}, {Lee}, {Lehmann}, {Lenon}, {Leonardi}, {Leroy},
  {Letendre}, {Levin}, {Li}, {Linker}, {Littenberg}, {Liu}, {Lo}, {Lockerbie}, {London}, {Lord}, {Lorenzini}, {Loriette}, {Lormand}, {Losurdo}, {Lough}, {Lousto}, {Lovelace}, {L{\"u}ck}, {Lumaca}, {Lundgren}, {Lynch}, {Ma}, {Macas}, {Macfoy}, {Machenschalk}, {MacInnis}, {Macleod}, {Maga{\~n}a Hernandez}, {Maga{\~n}a-Sandoval}, {Maga{\~n}a Zertuche}, {Magee}, {Majorana}, {Maksimovic}, {Man}, {Mandic}, {Mangano}, {Mansell}, {Manske}, {Mantovani}, {Marchesoni}, {Marion}, {M{\'a}rka}, {M{\'a}rka}, {Markakis}, {Markosyan}, {Markowitz}, {Maros}, {Marquina}, {Marsh}, {Martelli}, {Martellini}, {Martin}, {Martin}, {Martynov}, {Mason}, {Massera}, {Masserot}, {Massinger}, {Masso-Reid}, {Mastrogiovanni}, {Matas}, {Matichard}, {Matone}, {Mavalvala}, {Mazumder}, {McCarthy}, {McClelland}, {McCormick}, {McCuller}, {McGuire}, {McIntyre}, {McIver}, {McManus}, {McNeill}, {McRae}, {McWilliams}, {Meacher}, {Meadors}, {Mehmet}, {Meidam}, {Mejuto-Villa}, {Melatos}, {Mendell}, {Mercer}, {Merilh}, {Merzougui}, {Meshkov}, {Messenger},
  {Messick}, {Metzdorff}, {Meyers}, {Miao}, {Michel}, {Middleton}, {Mikhailov}, {Milano}, {Miller}, {Miller}, {Miller}, {Millhouse}, {Milovich-Goff}, {Minazzoli}, {Minenkov}, {Ming}, {Mishra}, {Mitra}, {Mitrofanov}, {Mitselmakher}, {Mittleman}, {Moffa}, {Moggi}, {Mogushi}, {Mohan}, {Mohapatra}, {Montani}, {Moore}, {Moraru}, {Moreno}, {Morriss}, {Mours}, {Mow-Lowry}, {Mueller}, {Muir}, {Mukherjee}, {Mukherjee}, {Mukherjee}, {Mukund}, {Mullavey}, {Munch}, {Mu{\~n}iz}, {Muratore}, {Murray}, {Napier}, {Nardecchia}, {Naticchioni}, {Nayak}, {Neilson}, {Nelemans}, {Nelson}, {Nery}, {Neunzert}, {Nevin}, {Newport}, {Newton}, {Ng}, {Nguyen}, {Nguyen}, {Nichols}, {Nielsen}, {Nissanke}, {Nitz}, {Noack}, {Nocera}, {Nolting}, {North}, {Nuttall}, {Oberling}, {O'Dea}, {Ogin}, {Oh}, {Oh}, {Ohme}, {Okada}, {Oliver}, {Oppermann}, {Oram}, {O'Reilly}, {Ormiston}, {Ortega}, {O'Shaughnessy}, {Ossokine}, {Ottaway}, {Overmier}, {Owen}, {Pace}, {Page}, {Page}, {Pai}, {Pai}, {Palamos}, {Palashov}, {Palomba}, {Pal-Singh}, {Pan}, {Pan},
  {Pang}, {Pang}, {Pankow}, {Pannarale}, {Pant}, {Paoletti}, {Paoli}, {Papa}, {Parida}, {Parker}, {Pascucci}, {Pasqualetti}, {Passaquieti}, {Passuello}, {Patil}, {Patricelli}, {Pearlstone}, {Pedraza}, {Pedurand}, {Pekowsky}, {Pele}, {Penn}, {Perez}, {Perreca}, {Perri}, {Pfeiffer}, {Phelps}, {Piccinni}, {Pichot}, {Piergiovanni}, {Pierro}, {Pillant}, {Pinard}, {Pinto}, {Pirello}, {Pitkin}, {Poe}, {Poggiani}, {Popolizio}, {Porter}, {Post}, {Powell}, {Prasad}, {Pratt}, {Pratten}, {Predoi}, {Prestegard}, {Price}, {Prijatelj}, {Principe}, {Privitera}, {Prodi}, {Prokhorov}, {Puncken}, {Punturo}, {Puppo}, {P{\"u}rrer}, {Qi}, {Quetschke}, {Quintero}, {Quitzow-James}, {Raab}, {Rabeling}, {Radkins}, {Raffai}, {Raja}, {Rajan}, {Rajbhandari}, {Rakhmanov}, {Ramirez}, {Ramos-Buades}, {Rapagnani}, {Raymond}, {Razzano}, {Read}, {Regimbau}, {Rei}, {Reid}, {Reitze}, {Ren}, {Reyes}, {Ricci}, {Ricker}, {Rieger}, {Riles}, {Rizzo}, {Robertson}, {Robie}, {Robinet}, {Rocchi}, {Rolland}, {Rollins}, {Roma}, {Romano}, {Romel}, {Romie},
  {Rosi{\'n}ska}, {Ross}, {Rowan}, {R{\"u}diger}, {Ruggi}, {Rutins}, {Ryan}, {Sachdev}, {Sadecki}, {Sadeghian}, {Sakellariadou}, {Salconi}, {Saleem}, {Salemi}, {Samajdar}, {Sammut}, {Sampson}, {Sanchez}, {Sanchez}, {Sanchis-Gual}, {Sandberg}, {Sanders}, {Sassolas}, {Sathyaprakash}, {Saulson}, {Sauter}, {Savage}, {Sawadsky}, {Schale}, {Scheel}, {Scheuer}, {Schmidt}, {Schmidt}, {Schnabel}, {Schofield}, {Sch{\"o}nbeck}, {Schreiber}, {Schuette}, {Schulte}, {Schutz}, {Schwalbe}, {Scott}, {Scott}, {Seidel}, {Sellers}, {Sengupta}, {Sentenac}, {Sequino}, {Sergeev}, {Shaddock}, {Shaffer}, {Shah}, {Shahriar}, {Shaner}, {Shao}, {Shapiro}, {Shawhan}, {Sheperd}, {Shoemaker}, {Shoemaker}, {Siellez}, {Siemens}, {Sieniawska}, {Sigg}, {Silva}, {Singer}, {Singh}, {Singhal}, {Sintes}, {Slagmolen}, {Smith}, {Smith}, {Smith}, {Somala}, {Son}, {Sonnenberg}, {Sorazu}, {Sorrentino}, {Souradeep}, {Spencer}, {Srivastava}, {Staats}, {Staley}, {Steinke}, {Steinlechner}, {Steinlechner}, {Steinmeyer}, {Stevenson}, {Stone}, {Stops},
  {Strain}, {Stratta}, {Strigin}, {Strunk}, {Sturani}, {Stuver}, {Summerscales}, {Sun}, {Sunil}, {Suresh}, {Sutton}, {Swinkels}, {Szczepa{\'n}czyk}, {Tacca}, {Tait}, {Talbot}, {Talukder}, {Tanner}, {T{\'a}pai}, {Taracchini}, {Tasson}, {Taylor}, {Taylor}, {Tewari}, {Theeg}, {Thies}, {Thomas}, {Thomas}, {Thomas}, {Thorne}, {Thorne}, {Thrane}, {Tiwari}, {Tiwari}, {Tokmakov}, {Toland}, {Tonelli}, {Tornasi}, {Torres-Forn{\'e}}, {Torrie}, {T{\"o}yr{\"a}}, {Travasso}, {Traylor}, {Trinastic}, {Tringali}, {Trozzo}, {Tsang}, {Tse}, {Tso}, {Tsukada}, {Tsuna}, {Tuyenbayev}, {Ueno}, {Ugolini}, {Unnikrishnan}, {Urban}, {Usman}, {Vahlbruch}, {Vajente}, {Valdes}, {van Bakel}, {van Beuzekom}, {van den Brand}, {Van Den Broeck}, {Vander-Hyde}, {van der Schaaf}, {van Heijningen}, {van Veggel}, {Vardaro}, {Varma}, {Vass}, {Vas{\'u}th}, {Vecchio}, {Vedovato}, {Veitch}, {Veitch}, {Venkateswara}, {Venugopalan}, {Verkindt}, {Vetrano}, {Vicer{\'e}}, {Viets}, {Vinciguerra}, {Vine}, {Vinet}, {Vitale}, {Vo}, {Vocca}, {Vorvick},
  {Vyatchanin}, {Wade}, {Wade}, {Wade}, {Walet}, {Walker}, {Wallace}, {Walsh}, {Wang}, {Wang}, {Wang}, {Wang}, {Wang}, {Ward}, {Warner}, {Was}, {Watchi}, {Weaver}, {Wei}, {Weinert}, {Weinstein}, {Weiss}, {Wen}, {Wessel}, {Wessels}, {Westerweck}, {Westphal}, {Wette}, {Whelan}, {Whitcomb}, {Whiting}, {Whittle}, {Wilken}, {Williams}, {Williams}, {Williamson}, {Willis}, {Willke}, {Wimmer}, {Winkler}, {Wipf}, {Wittel}, {Woan}, {Woehler}, {Wofford}, {Wong}, {Worden}, {Wright}, {Wu}, {Wysocki}, {Xiao}, {Yamamoto}, {Yancey}, {Yang}, {Yap}, {Yazback}, {Yu}, {Yu}, {Yvert}, {Zadro{\.z}ny}, {Zanolin}, {Zelenova}, {Zendri}, {Zevin}, {Zhang}, {Zhang}, {Zhang}, {Zhang}, {Zhao}, {Zhou}, {Zhou}, {Zhu}, {Zhu}, {Zimmerman}, {Zucker}, {Zweizig}, {LIGO Scientific Collaboration}, {Virgo Collaboration}, {Wilson-Hodge}, {Bissaldi}, {Blackburn}, {Briggs}, {Burns}, {Cleveland}, {Connaughton}, {Gibby}, {Giles}, {Goldstein}, {Hamburg}, {Jenke}, {Hui}, {Kippen}, {Kocevski}, {McBreen}, {Meegan}, {Paciesas}, {Poolakkil}, {Preece},
  {Racusin}, {Roberts}, {Stanbro}, {Veres}, {von Kienlin}, {GBM}, {Savchenko}, {Ferrigno}, {Kuulkers}, {Bazzano}, {Bozzo}, {Brandt}, {Chenevez}, {Courvoisier}, {Diehl}, {Domingo}, {Hanlon}, {Jourdain}, {Laurent}, {Lebrun}, {Lutovinov}, {Martin-Carrillo}, {Mereghetti}, {Natalucci}, {Rodi}, {Roques}, {Sunyaev}, {Ubertini}, {INTEGRAL}, {Aartsen}, {Ackermann}, {Adams}, {Aguilar}, {Ahlers}, {Ahrens}, {Samarai}, {Altmann}, {Andeen}, {Anderson}, {Ansseau}, {Anton}, {Arg{\"u}elles}, {Auffenberg}, {Axani}, {Bagherpour}, {Bai}, {Barron}, {Barwick}, {Baum}, {Bay}, {Beatty}, {Becker Tjus}, {Bernardini}, {Besson}, {Binder}, {Bindig}, {Blaufuss}, {Blot}, {Bohm}, {B{\"o}rner}, {Bos}, {Bose}, {B{\"o}ser}, {Botner}, {Bourbeau}, {Bourbeau}, {Bradascio}, {Braun}, {Brayeur}, {Brenzke}, {Bretz}, {Bron}, {Brostean-Kaiser}, {Burgman}, {Carver}, {Casey}, {Casier}, {Cheung}, {Chirkin}, {Christov}, {Clark}, {Classen}, {Coenders}, {Collin}, {Conrad}, {Cowen}, {Cross}, {Day}, {de Andr{\'e}}, {De Clercq}, {DeLaunay}, {Dembinski}, {De
  Ridder}, {Desiati}, {de Vries}, {de Wasseige}, {de With}, {DeYoung}, {D{\'\i}az-V{\'e}lez}, {di Lorenzo}, {Dujmovic}, {Dumm}, {Dunkman}, {Dvorak}, {Eberhardt}, {Ehrhardt}, {Eichmann}, {Eller}, {Evenson}, {Fahey}, {Fazely}, {Felde}, {Filimonov}, {Finley}, {Flis}, {Franckowiak}, {Friedman}, {Fuchs}, {Gaisser}, {Gallagher}, {Gerhardt}, {Ghorbani}, {Giang}, {Glauch}, {Gl{\"u}senkamp}, {Goldschmidt}, {Gonzalez}, {Grant}, {Griffith}, {Haack}, {Hallgren}, {Halzen}, {Hanson}, {Hebecker}, {Heereman}, {Helbing}, {Hellauer}, {Hickford}, {Hignight}, {Hill}, {Hoffman}, {Hoffmann}, {Hokanson-Fasig}, {Hoshina}, {Huang}, {Huber}, {Hultqvist}, {H{\"u}nnefeld}, {In}, {Ishihara}, {Jacobi}, {Japaridze}, {Jeong}, {Jero}, {Jones}, {Kalaczynski}, {Kang}, {Kappes}, {Karg}, {Karle}, {Kauer}, {Keivani}, {Kelley}, {Kheirandish}, {Kim}, {Kim}, {Kintscher}, {Kiryluk}, {Kittler}, {Klein}, {Kohnen}, {Koirala}, {Kolanoski}, {K{\"o}pke}, {Kopper}, {Kopper}, {Koschinsky}, {Koskinen}, {Kowalski}, {Krings}, {Kroll}, {Kr{\"u}ckl}, {Kunnen},
  {Kunwar}, {Kurahashi}, {Kuwabara}, {Kyriacou}, {Labare}, {Lanfranchi}, {Larson}, {Lauber}, {Lesiak-Bzdak}, {Leuermann}, {Liu}, {Lu}, {L{\"u}nemann}, {Luszczak}, {Madsen}, {Maggi}, {Mahn}, {Mancina}, {Maruyama}, {Mase}, {Maunu}, {McNally}, {Meagher}, {Medici}, {Meier}, {Menne}, {Merino}, {Meures}, {Miarecki}, {Micallef}, {Moment{\'e}}, {Montaruli}, {Moore}, {Moulai}, {Nahnhauer}, {Nakarmi}, {Naumann}, {Neer}, {Niederhausen}, {Nowicki}, {Nygren}, {Obertacke Pollmann}, {Olivas}, {O'Murchadha}, {Palczewski}, {Pandya}, {Pankova}, {Peiffer}, {Pepper}, {P{\'e}rez de los Heros}, {Pieloth}, {Pinat}, {Price}, {Przybylski}, {Raab}, {R{\"a}del}, {Rameez}, {Rawlins}, {Rea}, {Reimann}, {Relethford}, {Relich}, {Resconi}, {Rhode}, {Richman}, {Robertson}, {Rongen}, {Rott}, {Ruhe}, {Ryckbosch}, {Rysewyk}, {S{\"a}lzer}, {Sanchez Herrera}, {Sandrock}, {Sandroos}, {Santander}, {Sarkar}, {Sarkar}, {Satalecka}, {Schlunder}, {Schmidt}, {Schneider}, {Schoenen}, {Sch{\"o}neberg}, {Schumacher}, {Seckel}, {Seunarine}, {Soedingrekso},
  {Soldin}, {Song}, {Spiczak}, {Spiering}, {Stachurska}, {Stamatikos}, {Stanev}, {Stasik}, {Stettner}, {Steuer}, {Stezelberger}, {Stokstad}, {St{\"o}ssl}, {Strotjohann}, {Stuttard}, {Sullivan}, {Sutherland}, {Taboada}, {Tatar}, {Tenholt}, {Ter-Antonyan}, {Terliuk}, {Te{\v{s}}i{\'c}}, {Tilav}, {Toale}, {Tobin}, {Toscano}, {Tosi}, {Tselengidou}, {Tung}, {Turcati}, {Turley}, {Ty}, {Unger}, {Usner}, {Vandenbroucke}, {Van Driessche}, {van Eijndhoven}, {Vanheule}, {van Santen}, {Vehring}, {Vogel}, {Vraeghe}, {Walck}, {Wallace}, {Wallraff}, {Wandler}, {Wandkowsky}, {Waza}, {Weaver}, {Weiss}, {Wendt}, {Werthebach}, {Whelan}, {Wiebe}, {Wiebusch}, {Wille}, {Williams}, {Wills}, {Wolf}, {Wood}, {Woolsey}, {Woschnagg}, {Xu}, {Xu}, {Xu}, {Yanez}, {Yodh}, {Yoshida}, {Yuan}, {Zoll}, {IceCube Collaboration}, {Balasubramanian}, {Mate}, {Bhalerao}, {Bhattacharya}, {Vibhute}, {Dewangan}, {Rao}, {Vadawale}, {AstroSat Cadmium Zinc Telluride Imager Team}, {Svinkin}, {Hurley}, {Aptekar}, {Frederiks}, {Golenetskii}, {Kozlova},
  {Lysenko}, {Oleynik}, {Tsvetkova}, {Ulanov}, {Cline}, {IPN Collaboration}, {Li}, {Xiong}, {Zhang}, {Lu}, {Song}, {Cao}, {Chang}, {Chen}, {Chen}, {Chen}, {Chen}, {Chen}, {Chen}, {Cui}, {Cui}, {Deng}, {Dong}, {Du}, {Fu}, {Gao}, {Gao}, {Gao}, {Ge}, {Gu}, {Guan}, {Guo}, {Han}, {Hu}, {Huang}, {Huo}, {Jia}, {Jiang}, {Jiang}, {Jin}, {Jin}, {Li}, {Li}, {Li}, {Li}, {Li}, {Li}, {Li}, {Li}, {Li}, {Li}, {Li}, {Liang}, {Liao}, {Liu}, {Liu}, {Liu}, {Liu}, {Liu}, {Liu}, {Liu}, {Lu}, {Lu}, {Luo}, {Ma}, {Meng}, {Nang}, {Nie}, {Ou}, {Qu}, {Sai}, {Sun}, {Tan}, {Tao}, {Tao}, {Tuo}, {Wang}, {Wang}, {Wang}, {Wang}, {Wang}, {Wen}, {Wu}, {Wu}, {Xiao}, {Xu}, {Xu}, {Yan}, {Yang}, {Yang}, {Yang}, {Zhang}, {Zhang}, {Zhang}, {Zhang}, {Zhang}, {Zhang}, {Zhang}, {Zhang}, {Zhang}, {Zhang}, {Zhang}, {Zhang}, {Zhang}, {Zhang}, {Zhang}, {Zhang}, {Zhang}, {Zhang}, {Zhao}, {Zhao}, {Zhao}, {Zheng}, {Zhu}, {Zhu}, {Zou}, {Insight-HXMT Collaboration}, {Albert}, {Andr{\'e}}, {Anghinolfi}, {Ardid}, {Aubert}, {Aublin}, {Avgitas}, {Baret},
  {Barrios-Mart{\'\i}}, {Basa}, {Belhorma}, {Bertin}, {Biagi}, {Bormuth}, {Bourret}, {Bouwhuis}, {Br{\^a}nza{\c{s}}}, {Bruijn}, {Brunner}, {Busto}, {Capone}, {Caramete}, {Carr}, {Celli}, {Cherkaoui El Moursli}, {Chiarusi}, {Circella}, {Coelho}, {Coleiro}, {Coniglione}, {Costantini}, {Coyle}, {Creusot}, {D{\'\i}az}, {Deschamps}, {De Bonis}, {Distefano}, {Di Palma}, {Domi}, {Donzaud}, {Dornic}, {Drouhin}, {Eberl}, {El Bojaddaini}, {El Khayati}, {Els{\"a}sser}, {Enzenh{\"o}fer}, {Ettahiri}, {Fassi}, {Felis}, {Fusco}, {Gay}, {Giordano}, {Glotin}, {Gr{\'e}goire}, {Ruiz}, {Graf}, {Hallmann}, {van Haren}, {Heijboer}, {Hello}, {Hern{\'a}ndez-Rey}, {H{\"o}ssl}, {Hofest{\"a}dt}, {Hugon}, {Illuminati}, {James}, {de Jong}, {Jongen}, {Kadler}, {Kalekin}, {Katz}, {Kiessling}, {Kouchner}, {Kreter}, {Kreykenbohm}, {Kulikovskiy}, {Lachaud}, {Lahmann}, {Lef{\`e}vre}, {Leonora}, {Lotze}, {Loucatos}, {Marcelin}, {Margiotta}, {Marinelli}, {Mart{\'\i}nez-Mora}, {Mele}, {Melis}, {Michael}, {Migliozzi}, {Moussa}, {Navas}, {Nezri},
  {Organokov}, {P{\u{a}}v{\u{a}}la{\c{s}}}, {Pellegrino}, {Perrina}, {Piattelli}, {Popa}, {Pradier}, {Quinn}, {Racca}, {Riccobene}, {S{\'a}nchez-Losa}, {Salda{\~n}a}, {Salvadori}, {Samtleben}, {Sanguineti}, {Sapienza}, {Sieger}, {Spurio}, {Stolarczyk}, {Taiuti}, {Tayalati}, {Trovato}, {Turpin}, {T{\"o}nnis}, {Vallage}, {Van Elewyck}, {Versari}, {Vivolo}, {Vizzoca}, {Wilms}, {Zornoza}, {Z{\'u}{\~n}iga}, {ANTARES Collaboration}, {Beardmore}, {Breeveld}, {Burrows}, {Cenko}, {Cusumano}, {D'A{\`\i}}, {de Pasquale}, {Emery}, {Evans}, {Giommi}, {Gronwall}, {Kennea}, {Krimm}, {Kuin}, {Lien}, {Marshall}, {Melandri}, {Nousek}, {Oates}, {Osborne}, {Pagani}, {Page}, {Palmer}, {Perri}, {Siegel}, {Sbarufatti}, {Tagliaferri}, {Tohuvavohu}, {Swift Collaboration}, {Tavani}, {Verrecchia}, {Bulgarelli}, {Evangelista}, {Pacciani}, {Feroci}, {Pittori}, {Giuliani}, {Del Monte}, {Donnarumma}, {Argan}, {Trois}, {Ursi}, {Cardillo}, {Piano}, {Longo}, {Lucarelli}, {Munar-Adrover}, {Fuschino}, {Labanti}, {Marisaldi}, {Minervini},
  {Fioretti}, {Parmiggiani}, {Gianotti}, {Trifoglio}, {Di Persio}, {Antonelli}, {Barbiellini}, {Caraveo}, {Cattaneo}, {Costa}, {Colafrancesco}, {D'Amico}, {Ferrari}, {Morselli}, {Paoletti}, {Picozza}, {Pilia}, {Rappoldi}, {Soffitta}, {Vercellone}, {AGILE Team}, {Foley}, {Coulter}, {Kilpatrick}, {Drout}, {Piro}, {Shappee}, {Siebert}, {Simon}, {Ulloa}, {Kasen}, {Madore}, {Murguia-Berthier}, {Pan}, {Prochaska}, {Ramirez-Ruiz}, {Rest}, {Rojas-Bravo}, {1M2H Team}, {Berger}, {Soares-Santos}, {Annis}, {Alexander}, {Allam}, {Balbinot}, {Blanchard}, {Brout}, {Butler}, {Chornock}, {Cook}, {Cowperthwaite}, {Diehl}, {Drlica-Wagner}, {Drout}, {Durret}, {Eftekhari}, {Finley}, {Fong}, {Frieman}, {Fryer}, {Garc{\'\i}a-Bellido}, {Gruendl}, {Hartley}, {Herner}, {Kessler}, {Lin}, {Lopes}, {Louren{\c{c}}o}, {Margutti}, {Marshall}, {Matheson}, {Medina}, {Metzger}, {Mu{\~n}oz}, {Muir}, {Nicholl}, {Nugent}, {Palmese}, {Paz-Chinch{\'o}n}, {Quataert}, {Sako}, {Sauseda}, {Schlegel}, {Scolnic}, {Secco}, {Smith}, {Sobreira}, {Villar},
  {Vivas}, {Wester}, {Williams}, {Yanny}, {Zenteno}, {Zhang}, {Abbott}, {Banerji}, {Bechtol}, {Benoit-L{\'e}vy}, {Bertin}, {Brooks}, {Buckley-Geer}, {Burke}, {Capozzi}, {Carnero Rosell}, {Carrasco Kind}, {Castander}, {Crocce}, {Cunha}, {D'Andrea}, {da Costa}, {Davis}, {DePoy}, {Desai}, {Dietrich}, {Eifler}, {Fernandez}, {Flaugher}, {Fosalba}, {Gaztanaga}, {Gerdes}, {Giannantonio}, {Goldstein}, {Gruen}, {Gschwend}, {Gutierrez}, {Honscheid}, {James}, {Jeltema}, {Johnson}, {Johnson}, {Kent}, {Krause}, {Kron}, {Kuehn}, {Lahav}, {Lima}, {Maia}, {March}, {Martini}, {McMahon}, {Menanteau}, {Miller}, {Miquel}, {Mohr}, {Nichol}, {Ogando}, {Plazas}, {Romer}, {Roodman}, {Rykoff}, {Sanchez}, {Scarpine}, {Schindler}, {Schubnell}, {Sevilla-Noarbe}, {Sheldon}, {Smith}, {Smith}, {Stebbins}, {Suchyta}, {Swanson}, {Tarle}, {Thomas}, {Troxel}, {Tucker}, {Vikram}, {Walker}, {Wechsler}, {Weller}, {Carlin}, {Gill}, {Li}, {Marriner}, {Neilsen}, {Dark Energy Camera GW-EM Collaboration}, {DES Collaboration}, {Haislip}, {Kouprianov},
  {Reichart}, {Sand}, {Tartaglia}, {Valenti}, {Yang}, {DLT40 Collaboration}, {Benetti}, {Brocato}, {Campana}, {Cappellaro}, {Covino}, {D'Avanzo}, {D'Elia}, {Getman}, {Ghirlanda}, {Ghisellini}, {Limatola}, {Nicastro}, {Palazzi}, {Pian}, {Piranomonte}, {Possenti}, {Rossi}, {Salafia}, {Tomasella}, {Amati}, {Antonelli}, {Bernardini}, {Bufano}, {Capaccioli}, {Casella}, {Dadina}, {De Cesare}, {Di Paola}, {Giuffrida}, {Giunta}, {Israel}, {Lisi}, {Maiorano}, {Mapelli}, {Masetti}, {Pescalli}, {Pulone}, {Salvaterra}, {Schipani}, {Spera}, {Stamerra}, {Stella}, {Testa}, {Turatto}, {Vergani}, {Aresu}, {Bachetti}, {Buffa}, {Burgay}, {Buttu}, {Caria}, {Carretti}, {Casasola}, {Castangia}, {Carboni}, {Casu}, {Concu}, {Corongiu}, {Deiana}, {Egron}, {Fara}, {Gaudiomonte}, {Gusai}, {Ladu}, {Loru}, {Leurini}, {Marongiu}, {Melis}, {Melis}, {Migoni}, {Milia}, {Navarrini}, {Orlati}, {Ortu}, {Palmas}, {Pellizzoni}, {Perrodin}, {Pisanu}, {Poppi}, {Righini}, {Saba}, {Serra}, {Serrau}, {Stagni}, {Surcis}, {Vacca}, {Vargiu}, {Hunt},
  {Jin}, {Klose}, {Kouveliotou}, {Mazzali}, {M{\o}ller}, {Nava}, {Piran}, {Selsing}, {Vergani}, {Wiersema}, {Toma}, {Higgins}, {Mundell}, {di Serego Alighieri}, {G{\'o}tz}, {Gao}, {Gomboc}, {Kaper}, {Kobayashi}, {Kopac}, {Mao}, {Starling}, {Steele}, {van der Horst}, {GRAWITA: GRAvitational Wave Inaf TeAm}, {Acero}, {Atwood}, {Baldini}, {Barbiellini}, {Bastieri}, {Berenji}, {Bellazzini}, {Bissaldi}, {Blandford}, {Bloom}, {Bonino}, {Bottacini}, {Bregeon}, {Buehler}, {Buson}, {Cameron}, {Caputo}, {Caraveo}, {Cavazzuti}, {Chekhtman}, {Cheung}, {Chiang}, {Ciprini}, {Cohen-Tanugi}, {Cominsky}, {Costantin}, {Cuoco}, {D'Ammando}, {de Palma}, {Digel}, {Di Lalla}, {Di Mauro}, {Di Venere}, {Dubois}, {Fegan}, {Focke}, {Franckowiak}, {Fukazawa}, {Funk}, {Fusco}, {Gargano}, {Gasparrini}, {Giglietto}, {Giordano}, {Giroletti}, {Glanzman}, {Green}, {Grondin}, {Guillemot}, {Guiriec}, {Harding}, {Horan}, {J{\'o}hannesson}, {Kamae}, {Kensei}, {Kuss}, {La Mura}, {Latronico}, {Lemoine-Goumard}, {Longo}, {Loparco}, {Lovellette},
  {Lubrano}, {Magill}, {Maldera}, {Manfreda}, {Mazziotta}, {McEnery}, {Meyer}, {Michelson}, {Mirabal}, {Monzani}, {Moretti}, {Morselli}, {Moskalenko}, {Negro}, {Nuss}, {Ojha}, {Omodei}, {Orienti}, {Orlando}, {Palatiello}, {Paliya}, {Paneque}, {Pesce-Rollins}, {Piron}, {Porter}, {Principe}, {Rain{\`o}}, {Rando}, {Razzano}, {Razzaque}, {Reimer}, {Reimer}, {Reposeur}, {Rochester}, {Saz Parkinson}, {Sgr{\`o}}, {Siskind}, {Spada}, {Spandre}, {Suson}, {Takahashi}, {Tanaka}, {Thayer}, {Thayer}, {Thompson}, {Tibaldo}, {Torres}, {Torresi}, {Troja}, {Venters}, {Vianello}, {Zaharijas}, {Fermi Large Area Telescope Collaboration}, {Allison}, {Bannister}, {Dobie}, {Kaplan}, {Lenc}, {Lynch}, {Murphy}, {Sadler}, {Australia Telescope Compact Array}, {Hotan}, {James}, {Oslowski}, {Raja}, {Shannon}, {Whiting}, {Australian SKA Pathfinder}, {Arcavi}, {Howell}, {McCully}, {Hosseinzadeh}, {Hiramatsu}, {Poznanski}, {Barnes}, {Zaltzman}, {Vasylyev}, {Maoz}, {Las Cumbres Observatory Group}, {Cooke}, {Bailes}, {Wolf}, {Deller},
  {Lidman}, {Wang}, {Gendre}, {Andreoni}, {Ackley}, {Pritchard}, {Bessell}, {Chang}, {M{\"o}ller}, {Onken}, {Scalzo}, {Ridden-Harper}, {Sharp}, {Tucker}, {Farrell}, {Elmer}, {Johnston}, {Venkatraman Krishnan}, {Keane}, {Green}, {Jameson}, {Hu}, {Ma}, {Sun}, {Wu}, {Wang}, {Shang}, {Hu}, {Ashley}, {Yuan}, {Li}, {Tao}, {Zhu}, {Zhang}, {Suntzeff}, {Zhou}, {Yang}, {Orange}, {Morris}, {Cucchiara}, {Giblin}, {Klotz}, {Staff}, {Thierry}, {Schmidt}, {OzGrav}, {(Deeper}, {Wider}, {program}, {AST3}, {CAASTRO Collaborations}, {Tanvir}, {Levan}, {Cano}, {de Ugarte-Postigo}, {Gonz{\'a}lez-Fern{\'a}ndez}, {Greiner}, {Hjorth}, {Irwin}, {Kr{\"u}hler}, {Mandel}, {Milvang-Jensen}, {O'Brien}, {Rol}, {Rosetti}, {Rosswog}, {Rowlinson}, {Steeghs}, {Th{\"o}ne}, {Ulaczyk}, {Watson}, {Bruun}, {Cutter}, {Figuera Jaimes}, {Fujii}, {Fruchter}, {Gompertz}, {Jakobsson}, {Hodosan}, {J{\`e}rgensen}, {Kangas}, {Kann}, {Rabus}, {Schr{\o}der}, {Stanway}, {Wijers}, {VINROUGE Collaboration}, {Lipunov}, {Gorbovskoy}, {Kornilov}, {Tyurina},
  {Balanutsa}, {Kuznetsov}, {Vlasenko}, {Podesta}, {Lopez}, {Podesta}, {Levato}, {Saffe}, {Mallamaci}, {Budnev}, {Gress}, {Kuvshinov}, {Gorbunov}, {Vladimirov}, {Zimnukhov}, {Gabovich}, {Yurkov}, {Sergienko}, {Rebolo}, {Serra-Ricart}, {Tlatov}, {Ishmuhametova}, {MASTER Collaboration}, {Abe}, {Aoki}, {Aoki}, {Asakura}, {Baar}, {Barway}, {Bond}, {Doi}, {Finet}, {Fujiyoshi}, {Furusawa}, {Honda}, {Itoh}, {Kanda}, {Kawabata}, {Kawabata}, {Kim}, {Koshida}, {Kuroda}, {Lee}, {Liu}, {Matsubayashi}, {Miyazaki}, {Morihana}, {Morokuma}, {Motohara}, {Murata}, {Nagai}, {Nagashima}, {Nagayama}, {Nakaoka}, {Nakata}, {Ohsawa}, {Ohshima}, {Ohta}, {Okita}, {Saito}, {Saito}, {Sako}, {Sekiguchi}, {Sumi}, {Tajitsu}, {Takahashi}, {Takayama}, {Tamura}, {Tanaka}, {Tanaka}, {Terai}, {Tominaga}, {Tristram}, {Uemura}, {Utsumi}, {Yamaguchi}, {Yasuda}, {Yoshida}, {Zenko}, {J-GEM}, {Adams}, {Anupama}, {Bally}, {Barway}, {Bellm}, {Blagorodnova}, {Cannella}, {Chandra}, {Chatterjee}, {Clarke}, {Cobb}, {Cook}, {Copperwheat}, {De}, {Emery},
  {Feindt}, {Foster}, {Fox}, {Frail}, {Fremling}, {Frohmaier}, {Garcia}, {Ghosh}, {Giacintucci}, {Goobar}, {Gottlieb}, {Grefenstette}, {Hallinan}, {Harrison}, {Heida}, {Helou}, {Ho}, {Horesh}, {Hotokezaka}, {Ip}, {Itoh}, {Jacobs}, {Jencson}, {Kasen}, {Kasliwal}, {Kassim}, {Kim}, {Kiran}, {Kuin}, {Kulkarni}, {Kupfer}, {Lau}, {Madsen}, {Mazzali}, {Miller}, {Miyasaka}, {Mooley}, {Myers}, {Nakar}, {Ngeow}, {Nugent}, {Ofek}, {Palliyaguru}, {Pavana}, {Perley}, {Peters}, {Pike}, {Piran}, {Qi}, {Quimby}, {Rana}, {Rosswog}, {Rusu}, {Sadler}, {Van Sistine}, {Sollerman}, {Xu}, {Yan}, {Yatsu}, {Yu}, {Zhang}, {Zhao}, {GROWTH}, {JAGWAR}, {Caltech-NRAO}, {TTU-NRAO}, {NuSTAR Collaborations}, {Chambers}, {Huber}, {Schultz}, {Bulger}, {Flewelling}, {Magnier}, {Lowe}, {Wainscoat}, {Waters}, {Willman}, {Pan-STARRS}, {Ebisawa}, {Hanyu}, {Harita}, {Hashimoto}, {Hidaka}, {Hori}, {Ishikawa}, {Isobe}, {Iwakiri}, {Kawai}, {Kawai}, {Kawamuro}, {Kawase}, {Kitaoka}, {Makishima}, {Matsuoka}, {Mihara}, {Morita}, {Morita}, {Nakahira},
  {Nakajima}, {Nakamura}, {Negoro}, {Oda}, {Sakamaki}, {Sasaki}, {Serino}, {Shidatsu}, {Shimomukai}, {Sugawara}, {Sugita}, {Sugizaki}, {Tachibana}, {Takao}, {Tanimoto}, {Tomida}, {Tsuboi}, {Tsunemi}, {Ueda}, {Ueno}, {Yamada}, {Yamaoka}, {Yamauchi}, {Yatabe}, {Yoneyama}, {Yoshii}, {MAXI Team}, {Coward}, {Crisp}, {Macpherson}, {Andreoni}, {Laugier}, {Noysena}, {Klotz}, {Gendre}, {Thierry}, {Turpin}, {Consortium}, {Im}, {Choi}, {Kim}, {Yoon}, {Lim}, {Lee}, {Lee}, {Kim}, {Ko}, {Joe}, {Kwon}, {Kim}, {Lim}, {Choi}, {KU Collaboration}, {Fynbo}, {Malesani}, {Xu}, {Optical Telescope}, {Smartt}, {Jerkstrand}, {Kankare}, {Sim}, {Fraser}, {Inserra}, {Maguire}, {Leloudas}, {Magee}, {Shingles}, {Smith}, {Young}, {Kotak}, {Gal-Yam}, {Lyman}, {Homan}, {Agliozzo}, {Anderson}, {Angus}, {Ashall}, {Barbarino}, {Bauer}, {Berton}, {Botticella}, {Bulla}, {Cannizzaro}, {Cartier}, {Cikota}, {Clark}, {De Cia}, {Della Valle}, {Dennefeld}, {Dessart}, {Dimitriadis}, {Elias-Rosa}, {Firth}, {Fl{\"o}rs}, {Frohmaier}, {Galbany},
  {Gonz{\'a}lez-Gait{\'a}n}, {Gromadzki}, {Guti{\'e}rrez}, {Hamanowicz}, {Harmanen}, {Heintz}, {Hernandez}, {Hodgkin}, {Hook}, {Izzo}, {James}, {Jonker}, {Kerzendorf}, {Kostrzewa-Rutkowska}, {Kromer}, {Kuncarayakti}, {Lawrence}, {Manulis}, {Mattila}, {McBrien}, {M{\"u}ller}, {Nordin}, {O'Neill}, {Onori}, {Palmerio}, {Pastorello}, {Patat}, {Pignata}, {Podsiadlowski}, {Razza}, {Reynolds}, {Roy}, {Ruiter}, {Rybicki}, {Salmon}, {Pumo}, {Prentice}, {Seitenzahl}, {Smith}, {Sollerman}, {Sullivan}, {Szegedi}, {Taddia}, {Taubenberger}, {Terreran}, {Van Soelen}, {Vos}, {Walton}, {Wright}, {Wyrzykowski}, {Yaron}, {pre=''(''>ePESSTO}, {Chen}, {Kr{\"u}hler}, {Schady}, {Wiseman}, {Greiner}, {Rau}, {Schweyer}, {Klose}, {Nicuesa Guelbenzu}, {GROND}, {Palliyaguru}, {Tech University}, {Shara}, {Williams}, {Vaisanen}, {Potter}, {Romero Colmenero}, {Crawford}, {Buckley}, {Mao}, {SALT Group}, {D{\'\i}az}, {Macri}, {Garc{\'\i}a Lambas}, {Mendes de Oliveira}, {Nilo Castell{\'o}n}, {Ribeiro}, {S{\'a}nchez}, {Schoenell}, {Abramo},
  {Akras}, {Alcaniz}, {Artola}, {Beroiz}, {Bonoli}, {Cabral}, {Camuccio}, {Chavushyan}, {Coelho}, {Colazo}, {Costa-Duarte}, {Cuevas Larenas}, {Dom{\'\i}nguez Romero}, {Dultzin}, {Fern{\'a}ndez}, {Garc{\'\i}a}, {Girardini}, {Gon{\c{c}}alves}, {Gon{\c{c}}alves}, {Gurovich}, {Jim{\'e}nez-Teja}, {Kanaan}, {Lares}, {Lopes de Oliveira}, {L{\'o}pez-Cruz}, {Melia}, {Molino}, {Padilla}, {Pe{\~n}uela}, {Placco}, {Qui{\~n}ones}, {Ram{\'\i}rez Rivera}, {Renzi}, {Riguccini}, {R{\'\i}os-L{\'o}pez}, {Rodriguez}, {Sampedro}, {Schneiter}, {Sodr{\'e}}, {Starck}, {Torres-Flores}, {Tornatore}, {Zadro{\.z}ny}, {Castillo}, {TOROS: Transient Robotic Observatory of South Collaboration}, {Castro-Tirado}, {Tello}, {Hu}, {Zhang}, {Cunniffe}, {Castell{\'o}n}, {Hiriart}, {Caballero-Garc{\'\i}a}, {Jel{\'\i}nek}, {Kub{\'a}nek}, {P{\'e}rez del Pulgar}, {Park}, {Jeong}, {Castro Cer{\'o}n}, {Pandey}, {Yock}, {Querel}, {Fan}, {Wang}, {BOOTES Collaboration}, {Beardsley}, {Brown}, {Crosse}, {Emrich}, {Franzen}, {Gaensler}, {Horsley},
  {Johnston-Hollitt}, {Kenney}, {Morales}, {Pallot}, {Sokolowski}, {Steele}, {Tingay}, {Trott}, {Walker}, {Wayth}, {Williams}, {Wu}, {Murchison Widefield Array}, {Yoshida}, {Sakamoto}, {Kawakubo}, {Yamaoka}, {Takahashi}, {Asaoka}, {Ozawa}, {Torii}, {Shimizu}, {Tamura}, {Ishizaki}, {Cherry}, {Ricciarini}, {Penacchioni}, {Marrocchesi}, {CALET Collaboration}, {Pozanenko}, {Volnova}, {Mazaeva}, {Minaev}, {Krugov}, {Kusakin}, {Reva}, {Moskvitin}, {Rumyantsev}, {Inasaridze}, {Klunko}, {Tungalag}, {Schmalz}, {Burhonov}, {IKI-GW Follow-up Collaboration}, {Abdalla}, {Abramowski}, {Aharonian}, {Ait Benkhali}, {Ang{\"u}ner}, {Arakawa}, {Arrieta}, {Aubert}, {Backes}, {Balzer}, {Barnard}, {Becherini}, {Becker Tjus}, {Berge}, {Bernhard}, {Bernl{\"o}hr}, {Blackwell}, {B{\"o}ttcher}, {Boisson}, {Bolmont}, {Bonnefoy}, {Bordas}, {Bregeon}, {Brun}, {Brun}, {Bryan}, {B{\"u}chele}, {Bulik}, {Capasso}, {Caroff}, {Carosi}, {Casanova}, {Cerruti}, {Chakraborty}, {Chaves}, {Chen}, {Chevalier}, {Colafrancesco}, {Condon}, {Conrad},
  {Davids}, {Decock}, {Deil}, {Devin}, {deWilt}, {Dirson}, {Djannati-Ata{\"\i}}, {Donath}, {O'C. Drury}, {Dutson}, {Dyks}, {Edwards}, {Egberts}, {Emery}, {Ernenwein}, {Eschbach}, {Farnier}, {Fegan}, {Fernandes}, {Fiasson}, {Fontaine}, {Funk}, {F{\"u}ssling}, {Gabici}, {Gallant}, {Garrigoux}, {Gat{\'e}}, {Giavitto}, {Giebels}, {Glawion}, {Glicenstein}, {Gottschall}, {Grondin}, {Hahn}, {Haupt}, {Hawkes}, {Heinzelmann}, {Henri}, {Hermann}, {Hinton}, {Hofmann}, {Hoischen}, {Holch}, {Holler}, {Horns}, {Ivascenko}, {Iwasaki}, {Jacholkowska}, {Jamrozy}, {Jankowsky}, {Jankowsky}, {Jingo}, {Jouvin}, {Jung-Richardt}, {Kastendieck}, {Katarzy{\'n}ski}, {Katsuragawa}, {Kerszberg}, {Khangulyan}, {Kh{\'e}lifi}, {King}, {Klepser}, {Klochkov}, {Klu{\'z}niak}, {Komin}, {Kosack}, {Krakau}, {Kraus}, {Kr{\"u}ger}, {Laffon}, {Lamanna}, {Lau}, {Lees}, {Lefaucheur}, {Lemi{\`e}re}, {Lemoine-Goumard}, {Lenain}, {Leser}, {Lohse}, {Lorentz}, {Liu}, {Lypova}, {Malyshev}, {Marandon}, {Marcowith}, {Mariaud}, {Marx}, {Maurin}, {Maxted},
  {Mayer}, {Meintjes}, {Meyer}, {Mitchell}, {Moderski}, {Mohamed}, {Mohrmann}, {Mor{\r{a}}}, {Moulin}, {Murach}, {Nakashima}, {de Naurois}, {Ndiyavala}, {Niederwanger}, {Niemiec}, {Oakes}, {O'Brien}, {Odaka}, {Ohm}, {Ostrowski}, {Oya}, {Padovani}, {Panter}, {Parsons}, {Pekeur}, {Pelletier}, {Perennes}, {Petrucci}, {Peyaud}, {Piel}, {Pita}, {Poireau}, {Poon}, {Prokhorov}, {Prokoph}, {P{\"u}hlhofer}, {Punch}, {Quirrenbach}, {Raab}, {Rauth}, {Reimer}, {Reimer}, {Renaud}, {de los Reyes}, {Rieger}, {Rinchiuso}, {Romoli}, {Rowell}, {Rudak}, {Rulten}, {Sahakian}, {Saito}, {Sanchez}, {Santangelo}, {Sasaki}, {Schlickeiser}, {Sch{\"u}ssler}, {Schulz}, {Schwanke}, {Schwemmer}, {Seglar-Arroyo}, {Settimo}, {Seyffert}, {Shafi}, {Shilon}, {Shiningayamwe}, {Simoni}, {Sol}, {Spanier}, {Spir-Jacob}, {Stawarz}, {Steenkamp}, {Stegmann}, {Steppa}, {Sushch}, {Takahashi}, {Tavernet}, {Tavernier}, {Taylor}, {Terrier}, {Tibaldo}, {Tiziani}, {Tluczykont}, {Trichard}, {Tsirou}, {Tsuji}, {Tuffs}, {Uchiyama}, {van der Walt}, {van Eldik},
  {van Rensburg}, {van Soelen}, {Vasileiadis}, {Veh}, {Venter}, {Viana}, {Vincent}, {Vink}, {Voisin}, {V{\"o}lk}, {Vuillaume}, {Wadiasingh}, {Wagner}, {Wagner}, {Wagner}, {White}, {Wierzcholska}, {Willmann}, {W{\"o}rnlein}, {Wouters}, {Yang}, {Zaborov}, {Zacharias}, {Zanin}, {Zdziarski}, {Zech}, {Zefi}, {Ziegler}, {Zorn}, {{\.Z}ywucka}, {H.~E.~S.~S. Collaboration}, {Fender}, {Broderick}, {Rowlinson}, {Wijers}, {Stewart}, {ter Veen}, {Shulevski}, {LOFAR Collaboration}, {Kavic}, {Simonetti}, {League}, {Tsai}, {Obenberger}, {Nathaniel}, {Taylor}, {Dowell}, {Liebling}, {Estes}, {Lippert}, {Sharma}, {Vincent}, {Farella}, {Wavelength Array}, {Abeysekara}, {Albert}, {Alfaro}, {Alvarez}, {Arceo}, {Arteaga-Vel{\'a}zquez}, {Avila Rojas}, {Ayala Solares}, {Barber}, {Becerra Gonzalez}, {Becerril}, {Belmont-Moreno}, {BenZvi}, {Berley}, {Bernal}, {Braun}, {Brisbois}, {Caballero-Mora}, {Capistr{\'a}n}, {Carrami{\~n}ana}, {Casanova}, {Castillo}, {Cotti}, {Cotzomi}, {Couti{\~n}o de Le{\'o}n}, {De Le{\'o}n}, {De la Fuente},
  {Diaz Hernandez}, {Dichiara}, {Dingus}, {DuVernois}, {D{\'\i}az-V{\'e}lez}, {Ellsworth}, {Engel}, {Enr{\'\i}quez-Rivera}, {Fiorino}, {Fleischhack}, {Fraija}, {Garc{\'\i}a-Gonz{\'a}lez}, {Garfias}, {Gerhardt}, {Gonz{\~o}lez Mu{\~n}oz}, {Gonz{\'a}lez}, {Goodman}, {Hampel-Arias}, {Harding}, {Hernandez}, {Hernandez-Almada}, {Hona}, {H{\"u}ntemeyer}, {Iriarte}, {Jardin-Blicq}, {Joshi}, {Kaufmann}, {Kieda}, {Lara}, {Lauer}, {Lennarz}, {Le{\'o}n Vargas}, {Linnemann}, {Longinotti}, {Raya}, {Luna-Garc{\'\i}a}, {L{\'o}pez-Coto}, {Malone}, {Marinelli}, {Martinez}, {Martinez-Castellanos}, {Mart{\'\i}nez-Castro}, {Mart{\'\i}nez-Huerta}, {Matthews}, {Miranda-Romagnoli}, {Moreno}, {Mostaf{\'a}}, {Nellen}, {Newbold}, {Nisa}, {Noriega-Papaqui}, {Pelayo}, {Pretz}, {P{\'e}rez-P{\'e}rez}, {Ren}, {Rho}, {Rivi{\`e}re}, {Rosa-Gonz{\'a}lez}, {Rosenberg}, {Ruiz-Velasco}, {Salazar}, {Salesa Greus}, {Sandoval}, {Schneider}, {Schoorlemmer}, {Sinnis}, {Smith}, {Springer}, {Surajbali}, {Tibolla}, {Tollefson}, {Torres}, {Ukwatta},
  {Weisgarber}, {Westerhoff}, {Wisher}, {Wood}, {Yapici}, {Yodh}, {Younk}, {Zhou}, {{\'A}lvarez}, {HAWC Collaboration}, {Aab}, {Abreu}, {Aglietta}, {Albuquerque}, {Albury}, {Allekotte}, {Almela}, {Alvarez Castillo}, {Alvarez-Mu{\~n}iz}, {Anastasi}, {Anchordoqui}, {Andrada}, {Andringa}, {Aramo}, {Arsene}, {Asorey}, {Assis}, {Avila}, {Badescu}, {Balaceanu}, {Barbato}, {Barreira Luz}, {Becker}, {Bellido}, {Berat}, {Bertaina}, {Bertou}, {Biermann}, {Biteau}, {Blaess}, {Blanco}, {Blazek}, {Bleve}, {Boh{\'a}{\v{c}}ov{\'a}}, {Bonifazi}, {Borodai}, {Botti}, {Brack}, {Brancus}, {Bretz}, {Bridgeman}, {Briechle}, {Buchholz}, {Bueno}, {Buitink}, {Buscemi}, {Caballero-Mora}, {Caccianiga}, {Cancio}, {Canfora}, {Caruso}, {Castellina}, {Catalani}, {Cataldi}, {Cazon}, {Chavez}, {Chinellato}, {Chudoba}, {Clay}, {Cobos Cerutti}, {Colalillo}, {Coleman}, {Collica}, {Coluccia}, {Concei{\c{c}}{\~a}o}, {Consolati}, {Contreras}, {Cooper}, {Coutu}, {Covault}, {Cronin}, {D'Amico}, {Daniel}, {Dasso}, {Daumiller}, {Dawson}, {Day}, {de
  Almeida}, {de Jong}, {De Mauro}, {de Mello Neto}, {De Mitri}, {de Oliveira}, {de Souza}, {Debatin}, {Deligny}, {D{\'\i}az Castro}, {Diogo}, {Dobrigkeit}, {D'Olivo}, {Dorosti}, {Dos Anjos}, {Dova}, {Dundovic}, {Ebr}, {Engel}, {Erdmann}, {Erfani}, {Escobar}, {Espadanal}, {Etchegoyen}, {Falcke}, {Farmer}, {Farrar}, {Fauth}, {Fazzini}, {Feldbusch}, {Fenu}, {Fick}, {Figueira}, {Filip{\v{c}}i{\v{c}}}, {Freire}, {Fujii}, {Fuster}, {Ga{\"\i}or}, {Garc{\'\i}a}, {Gat{\'e}}, {Gemmeke}, {Gherghel-Lascu}, {Ghia}, {Giaccari}, {Giammarchi}, {Giller}, {G{\l}as}, {Glaser}, {Golup}, {G{\'o}mez Berisso}, {G{\'o}mez Vitale}, {Gonz{\'a}lez}, {Gorgi}, {Gottowik}, {Grillo}, {Grubb}, {Guarino}, {Guedes}, {Halliday}, {Hampel}, {Hansen}, {Harari}, {Harrison}, {Harvey}, {Haungs}, {Hebbeker}, {Heck}, {Heimann}, {Herve}, {Hill}, {Hojvat}, {Holt}, {Homola}, {H{\"o}randel}, {Horvath}, {Hrabovsk{\'y}}, {Huege}, {Hulsman}, {Insolia}, {Isar}, {Jandt}, {Johnsen}, {Josebachuili}, {Jurysek}, {K{\"a}{\"a}p{\"a}}, {Kampert}, {Keilhauer},
  {Kemmerich}, {Kemp}, {Kieckhafer}, {Klages}, {Kleifges}, {Kleinfeller}, {Krause}, {Krohm}, {Kuempel}, {Kukec Mezek}, {Kunka}, {Kuotb Awad}, {Lago}, {LaHurd}, {Lang}, {Lauscher}, {Legumina}, {Leigui de Oliveira}, {Letessier-Selvon}, {Lhenry-Yvon}, {Link}, {Lo Presti}, {Lopes}, {L{\'o}pez}, {L{\'o}pez Casado}, {Lorek}, {Luce}, {Lucero}, {Malacari}, {Mallamaci}, {Mandat}, {Mantsch}, {Mariazzi}, {Maris}, {Marsella}, {Martello}, {Martinez}, {Mart{\'\i}nez Bravo}, {Mas{\'\i}as Meza}, {Mathes}, {Mathys}, {Matthews}, {Matthiae}, {Mayotte}, {Mazur}, {Medina}, {Medina-Tanco}, {Melo}, {Menshikov}, {Merenda}, {Michal}, {Micheletti}, {Middendorf}, {Miramonti}, {Mitrica}, {Mockler}, {Mollerach}, {Montanet}, {Morello}, {Morlino}, {M{\"u}ller}, {M{\"u}ller}, {Muller}, {M{\"u}ller}, {Mussa}, {Naranjo}, {Nguyen}, {Niculescu-Oglinzanu}, {Niechciol}, {Niemietz}, {Niggemann}, {Nitz}, {Nosek}, {Novotny}, {No{\v{z}}ka}, {N{\'u}{\~n}ez}, {Oikonomou}, {Olinto}, {Palatka}, {Pallotta}, {Papenbreer}, {Parente}, {Parra}, {Paul},
  {Pech}, {Pedreira}, {P{\c{e}}kala}, {Pe{\~n}a-Rodriguez}, {Pereira}, {Perlin}, {Perrone}, {Peters}, {Petrera}, {Phuntsok}, {Pierog}, {Pimenta}, {Pirronello}, {Platino}, {Plum}, {Poh}, {Porowski}, {Prado}, {Privitera}, {Prouza}, {Quel}, {Querchfeld}, {Quinn}, {Ramos-Pollan}, {Rautenberg}, {Ravignani}, {Ridky}, {Riehn}, {Risse}, {Ristori}, {Rizi}, {Rodrigues de Carvalho}, {Rodriguez Fernandez}, {Rodriguez Rojo}, {Roncoroni}, {Roth}, {Roulet}, {Rovero}, {Ruehl}, {Saffi}, {Saftoiu}, {Salamida}, {Salazar}, {Saleh}, {Salina}, {S{\'a}nchez}, {Sanchez-Lucas}, {Santos}, {Santos}, {Sarazin}, {Sarmento}, {Sarmiento-Cano}, {Sato}, {Schauer}, {Scherini}, {Schieler}, {Schimp}, {Schmidt}, {Scholten}, {Schov{\'a}nek}, {Schr{\"o}der}, {Schr{\"o}der}, {Schulz}, {Schumacher}, {Sciutto}, {Segreto}, {Shadkam}, {Shellard}, {Sigl}, {Silli}, {{\v{S}}m{\'\i}da}, {Snow}, {Sommers}, {Sonntag}, {Soriano}, {Squartini}, {Stanca}, {Stani{\v{c}}}, {Stasielak}, {Stassi}, {Stolpovskiy}, {Strafella}, {Streich}, {Suarez}, {Suarez-Dur{\'a}n},
  {Sudholz}, {Suomij{\"a}rvi}, {Supanitsky}, {{\v{S}}up{\'\i}k}, {Swain}, {Szadkowski}, {Taboada}, {Taborda}, {Timmermans}, {Todero Peixoto}, {Tomankova}, {Tom{\'e}}, {Torralba Elipe}, {Travnicek}, {Trini}, {Tueros}, {Ulrich}, {Unger}, {Urban}, {Vald{\'e}s Galicia}, {Vali{\~n}o}, {Valore}, {van Aar}, {van Bodegom}, {van den Berg}, {van Vliet}, {Varela}, {Vargas C{\'a}rdenas}, {V{\'a}zquez}, {Veberi{\v{c}}}, {Ventura}, {Vergara Quispe}, {Verzi}, {Vicha}, {Villase{\~n}or}, {Vorobiov}, {Wahlberg}, {Wainberg}, {Walz}, {Watson}, {Weber}, {Weindl}, {Wiede{\'n}ski}, {Wiencke}, {Wilczy{\'n}ski}, {Wirtz}, {Wittkowski}, {Wundheiler}, {Yang}, {Yushkov}, {Zas}, {Zavrtanik}, {Zavrtanik}, {Zepeda}, {Zimmermann}, {Ziolkowski}, {Zong}, {Zuccarello}, {Pierre Auger Collaboration}, {Kim}, {Schulze}, {Bauer}, {Corral-Santana}, {de Gregorio-Monsalvo}, {Gonz{\'a}lez-L{\'o}pez}, {Hartmann}, {Ishwara-Chandra}, {Mart{\'\i}n}, {Mehner}, {Misra}, {Micha{\l}owski}, {Resmi}, {ALMA Collaboration}, {Paragi}, {Agudo}, {An}, {Beswick},
  {Casadio}, {Frey}, {Jonker}, {Kettenis}, {Marcote}, {Moldon}, {Szomoru}, {van Langevelde}, {Yang}, {Euro VLBI Team}, {Cwiek}, {Cwiok}, {Czyrkowski}, {Dabrowski}, {Kasprowicz}, {Mankiewicz}, {Nawrocki}, {Opiela}, {Piotrowski}, {Wrochna}, {Zaremba}, {{\.Z}arnecki}, {Pi of Sky Collaboration}, {Haggard}, {Nynka}, {Ruan}, {Chandra Team at McGill University}, {Bland}, {Booler}, {Devillepoix}, {de Gois}, {Hancock}, {Howie}, {Paxman}, {Sansom}, {Towner}, {Desert Fireball Network}, {Tonry}, {Coughlin}, {Stubbs}, {Denneau}, {Heinze}, {Stalder}, {Weiland}, {ATLAS}, {Eatough}, {Kramer}, {Kraus}, {Time Resolution Universe Survey}, {Troja}, {Piro}, {Becerra Gonz{\'a}lez}, {Butler}, {Fox}, {Khandrika}, {Kutyrev}, {Lee}, {Ricci}, {Ryan}, {S{\'a}nchez-Ram{\'\i}rez}, {Veilleux}, {Watson}, {Wieringa}, {Burgess}, {van Eerten}, {Fontes}, {Fryer}, {Korobkin}, {Wollaeger}, {RIMAS}, {RATIR}, {Camilo}, {Foley}, {Goedhart}, {Makhathini}, {Oozeer}, {Smirnov}, {Fender}, {Woudt}, \& {South Africa/MeerKAT}}]{2017ApJ...848L..12A}
{Abbott}, B.~P., {Abbott}, R., {Abbott}, T.~D., {et~al.} 2017, \apjl, 848, L12

\bibitem[{{Arcones} \& {Mart{\'\i}nez-Pinedo}(2011)}]{2011PhRvC..83d5809A}
{Arcones}, A. \& {Mart{\'\i}nez-Pinedo}, G. 2011, \prc, 83, 045809

\bibitem[{{Arcones} \& {Thielemann}(2023)}]{2023A&ARv..31....1A}
{Arcones}, A. \& {Thielemann}, F.-K. 2023, \aapr, 31, 1

\bibitem[{{Arnould} {et~al.}(2007){Arnould}, {Goriely}, \& {Takahashi}}]{2007PhR...450...97A}
{Arnould}, M., {Goriely}, S., \& {Takahashi}, K. 2007, \physrep, 450, 97

\bibitem[{{Burbidge} {et~al.}(1957){Burbidge}, {Burbidge}, {Fowler}, \& {Hoyle}}]{1957RvMP...29..547B}
{Burbidge}, E.~M., {Burbidge}, G.~R., {Fowler}, W.~A., \& {Hoyle}, F. 1957, Reviews of Modern Physics, 29, 547

\bibitem[{{Chen} {et~al.}(2022){Chen}, {Hu}, \& {Liang}}]{2022ApJ...932L...7C}
{Chen}, M.-H., {Hu}, R.-C., \& {Liang}, E.-W. 2022, \apjl, 932, L7

\bibitem[{{Chen} {et~al.}(2023){Chen}, {Hu}, \& {Liang}}]{2023MNRAS.520.2806C}
{Chen}, M.-H., {Hu}, R.-C., \& {Liang}, E.-W. 2023, \mnras, 520, 2806

\bibitem[{{Chen} {et~al.}(2024{\natexlab{a}}){Chen}, {Li}, {Chen}, {Hu}, \& {Liang}}]{2024MNRAS.529.1154C}
{Chen}, M.-H., {Li}, L.-X., {Chen}, Q.-H., {Hu}, R.-C., \& {Liang}, E.-W. 2024{\natexlab{a}}, \mnras, 529, 1154

\bibitem[{{Chen} {et~al.}(2024{\natexlab{b}}){Chen}, {Li}, \& {Liang}}]{2024ApJ...971..143C}
{Chen}, M.-H., {Li}, L.-X., \& {Liang}, E.-W. 2024{\natexlab{b}}, \apj, 971, 143

\bibitem[{{Chen} {et~al.}(2021){Chen}, {Li}, {Lin}, \& {Liang}}]{2021ApJ...919...59C}
{Chen}, M.-H., {Li}, L.-X., {Lin}, D.-B., \& {Liang}, E.-W. 2021, \apj, 919, 59

\bibitem[{{Chen} \& {Liang}(2024)}]{2024MNRAS.527.5540C}
{Chen}, M.-H. \& {Liang}, E.-W. 2024, \mnras, 527, 5540

\bibitem[{{Cowan} {et~al.}(2002){Cowan}, {Sneden}, {Burles}, {Ivans}, {Beers}, {Truran}, {Lawler}, {Primas}, {Fuller}, {Pfeiffer}, \& {Kratz}}]{2002ApJ...572..861C}
{Cowan}, J.~J., {Sneden}, C., {Burles}, S., {et~al.} 2002, \apj, 572, 861

\bibitem[{{Cowan} {et~al.}(2021){Cowan}, {Sneden}, {Lawler}, {Aprahamian}, {Wiescher}, {Langanke}, {Mart{\'\i}nez-Pinedo}, \& {Thielemann}}]{2021RvMP...93a5002C}
{Cowan}, J.~J., {Sneden}, C., {Lawler}, J.~E., {et~al.} 2021, Reviews of Modern Physics, 93, 015002

\bibitem[{{Cyburt} {et~al.}(2010){Cyburt}, {Amthor}, {Ferguson}, {Meisel}, {Smith}, {Warren}, {Heger}, {Hoffman}, {Rauscher}, {Sakharuk}, {Schatz}, {Thielemann}, \& {Wiescher}}]{2010ApJS..189..240C}
{Cyburt}, R.~H., {Amthor}, A.~M., {Ferguson}, R., {et~al.} 2010, \apjs, 189, 240

\bibitem[{{Duflo} \& {Zuker}(1995)}]{1995PhRvC..52...23D}
{Duflo}, J. \& {Zuker}, A.~P. 1995, \prc, 52, R23

\bibitem[{{Eichler} {et~al.}(2015){Eichler}, {Arcones}, {Kelic}, {Korobkin}, {Langanke}, {Marketin}, {Martinez-Pinedo}, {Panov}, {Rauscher}, {Rosswog}, {Winteler}, {Zinner}, \& {Thielemann}}]{2015ApJ...808...30E}
{Eichler}, M., {Arcones}, A., {Kelic}, A., {et~al.} 2015, \apj, 808, 30

\bibitem[{{Frebel} {et~al.}(2007){Frebel}, {Christlieb}, {Norris}, {Thom}, {Beers}, \& {Rhee}}]{2007ApJ...660L.117F}
{Frebel}, A., {Christlieb}, N., {Norris}, J.~E., {et~al.} 2007, \apjl, 660, L117

\bibitem[{{Goriely} {et~al.}(2009){Goriely}, {Chamel}, \& {Pearson}}]{2009PhRvL.102o2503G}
{Goriely}, S., {Chamel}, N., \& {Pearson}, J.~M. 2009, \prl, 102, 152503

\bibitem[{{Goriely} {et~al.}(2013){Goriely}, {Chamel}, \& {Pearson}}]{2013PhRvC..88f1302G}
{Goriely}, S., {Chamel}, N., \& {Pearson}, J.~M. 2013, \prc, 88, 061302

\bibitem[{{Goriely} {et~al.}(2008){Goriely}, {Hilaire}, \& {Koning}}]{2008A&A...487..767G}
{Goriely}, S., {Hilaire}, S., \& {Koning}, A.~J. 2008, \aap, 487, 767

\bibitem[{{Hao} {et~al.}(2023){Hao}, {Niu}, \& {Niu}}]{2023PhLB..84438092H}
{Hao}, Y.~W., {Niu}, Y.~F., \& {Niu}, Z.~M. 2023, Physics Letters B, 844, 138092

\bibitem[{{Hill} {et~al.}(2017){Hill}, {Christlieb}, {Beers}, {Barklem}, {Kratz}, {Nordstr{\"o}m}, {Pfeiffer}, \& {Farouqi}}]{2017A&A...607A..91H}
{Hill}, V., {Christlieb}, N., {Beers}, T.~C., {et~al.} 2017, \aap, 607, A91

\bibitem[{{Hill} {et~al.}(2002){Hill}, {Plez}, {Cayrel}, {Beers}, {Nordstr{\"o}m}, {Andersen}, {Spite}, {Spite}, {Barbuy}, {Bonifacio}, {Depagne}, {Fran{\c{c}}ois}, \& {Primas}}]{2002A&A...387..560H}
{Hill}, V., {Plez}, B., {Cayrel}, R., {et~al.} 2002, \aap, 387, 560

\bibitem[{{Holmbeck} {et~al.}(2018){Holmbeck}, {Beers}, {Roederer}, {Placco}, {Hansen}, {Sakari}, {Sneden}, {Liu}, {Lee}, {Cowan}, \& {Frebel}}]{2018ApJ...859L..24H}
{Holmbeck}, E.~M., {Beers}, T.~C., {Roederer}, I.~U., {et~al.} 2018, \apjl, 859, L24

\bibitem[{{Horowitz} {et~al.}(2019){Horowitz}, {Arcones}, {C{\^o}t{\'e}}, {Dillmann}, {Nazarewicz}, {Roederer}, {Schatz}, {Aprahamian}, {Atanasov}, {Bauswein}, {Beers}, {Bliss}, {Brodeur}, {Clark}, {Frebel}, {Foucart}, {Hansen}, {Just}, {Kankainen}, {McLaughlin}, {Kelly}, {Liddick}, {Lee}, {Lippuner}, {Martin}, {Mendoza-Temis}, {Metzger}, {Mumpower}, {Perdikakis}, {Pereira}, {O'Shea}, {Reifarth}, {Rogers}, {Siegel}, {Spyrou}, {Surman}, {Tang}, {Uesaka}, \& {Wang}}]{2019JPhG...46h3001H}
{Horowitz}, C.~J., {Arcones}, A., {C{\^o}t{\'e}}, B., {et~al.} 2019, Journal of Physics G Nuclear Physics, 46, 083001

\bibitem[{{Hotokezaka} {et~al.}(2018){Hotokezaka}, {Beniamini}, \& {Piran}}]{2018IJMPD..2742005H}
{Hotokezaka}, K., {Beniamini}, P., \& {Piran}, T. 2018, International Journal of Modern Physics D, 27, 1842005

\bibitem[{{Hotokezaka} {et~al.}(2016){Hotokezaka}, {Wanajo}, {Tanaka}, {Bamba}, {Terada}, \& {Piran}}]{2016MNRAS.459...35H}
{Hotokezaka}, K., {Wanajo}, S., {Tanaka}, M., {et~al.} 2016, \mnras, 459, 35

\bibitem[{{Kajino} {et~al.}(2019){Kajino}, {Aoki}, {Balantekin}, {Diehl}, {Famiano}, \& {Mathews}}]{2019PrPNP.107..109K}
{Kajino}, T., {Aoki}, W., {Balantekin}, A.~B., {et~al.} 2019, Progress in Particle and Nuclear Physics, 107, 109

\bibitem[{{Kasen} {et~al.}(2017){Kasen}, {Metzger}, {Barnes}, {Quataert}, \& {Ramirez-Ruiz}}]{2017Natur.551...80K}
{Kasen}, D., {Metzger}, B., {Barnes}, J., {Quataert}, E., \& {Ramirez-Ruiz}, E. 2017, \nat, 551, 80

\bibitem[{{Kodama} \& {Takahashi}(1975)}]{1975NuPhA.239..489K}
{Kodama}, T. \& {Takahashi}, K. 1975, \nphysa, 239, 489

\bibitem[{{Kondev} {et~al.}(2021){Kondev}, {Wang}, {Huang}, {Naimi}, \& {Audi}}]{2021ChPhC..45c0001K}
{Kondev}, F.~G., {Wang}, M., {Huang}, W.~J., {Naimi}, S., \& {Audi}, G. 2021, Chinese Physics C, 45, 030001

\bibitem[{{Lattimer} \& {Schramm}(1974)}]{1974ApJ...192L.145L}
{Lattimer}, J.~M. \& {Schramm}, D.~N. 1974, \apjl, 192, L145

\bibitem[{{Levan} {et~al.}(2024){Levan}, {Gompertz}, {Salafia}, {Bulla}, {Burns}, {Hotokezaka}, {Izzo}, {Lamb}, {Malesani}, {Oates}, {Ravasio}, {Rouco Escorial}, {Schneider}, {Sarin}, {Schulze}, {Tanvir}, {Ackley}, {Anderson}, {Brammer}, {Christensen}, {Dhillon}, {Evans}, {Fausnaugh}, {Fong}, {Fruchter}, {Fryer}, {Fynbo}, {Gaspari}, {Heintz}, {Hjorth}, {Kennea}, {Kennedy}, {Laskar}, {Leloudas}, {Mandel}, {Martin-Carrillo}, {Metzger}, {Nicholl}, {Nugent}, {Palmerio}, {Pugliese}, {Rastinejad}, {Rhodes}, {Rossi}, {Saccardi}, {Smartt}, {Stevance}, {Tohuvavohu}, {van der Horst}, {Vergani}, {Watson}, {Barclay}, {Bhirombhakdi}, {Breedt}, {Breeveld}, {Brown}, {Campana}, {Chrimes}, {D'Avanzo}, {D'Elia}, {De Pasquale}, {Dyer}, {Galloway}, {Garbutt}, {Green}, {Hartmann}, {Jakobsson}, {Kerry}, {Kouveliotou}, {Langeroodi}, {Le Floc'h}, {Leung}, {Littlefair}, {Munday}, {O'Brien}, {Parsons}, {Pelisoli}, {Sahman}, {Salvaterra}, {Sbarufatti}, {Steeghs}, {Tagliaferri}, {Th{\"o}ne}, {de Ugarte Postigo}, \&
  {Kann}}]{2024Natur.626..737L}
{Levan}, A.~J., {Gompertz}, B.~P., {Salafia}, O.~S., {et~al.} 2024, \nat, 626, 737

\bibitem[{{Li}(2019)}]{2019ApJ...872...19L}
{Li}, L.-X. 2019, \apj, 872, 19

\bibitem[{{Li} \& {Paczy{\'n}ski}(1998)}]{1998ApJ...507L..59L}
{Li}, L.-X. \& {Paczy{\'n}ski}, B. 1998, \apjl, 507, L59

\bibitem[{{Lippuner} \& {Roberts}(2015)}]{2015ApJ...815...82L}
{Lippuner}, J. \& {Roberts}, L.~F. 2015, \apj, 815, 82

\bibitem[{{Lippuner} \& {Roberts}(2017)}]{2017ApJS..233...18L}
{Lippuner}, J. \& {Roberts}, L.~F. 2017, \apjs, 233, 18

\bibitem[{{Mendoza-Temis} {et~al.}(2015){Mendoza-Temis}, {Wu}, {Langanke}, {Mart{\'\i}nez-Pinedo}, {Bauswein}, \& {Janka}}]{2015PhRvC..92e5805M}
{Mendoza-Temis}, J. d.~J., {Wu}, M.-R., {Langanke}, K., {et~al.} 2015, \prc, 92, 055805

\bibitem[{{Metzger} {et~al.}(2010){Metzger}, {Mart{\'\i}nez-Pinedo}, {Darbha}, {Quataert}, {Arcones}, {Kasen}, {Thomas}, {Nugent}, {Panov}, \& {Zinner}}]{2010MNRAS.406.2650M}
{Metzger}, B.~D., {Mart{\'\i}nez-Pinedo}, G., {Darbha}, S., {et~al.} 2010, \mnras, 406, 2650

\bibitem[{{M{\"o}ller} {et~al.}(2012){M{\"o}ller}, {Myers}, {Sagawa}, \& {Yoshida}}]{2012PhRvL.108e2501M}
{M{\"o}ller}, P., {Myers}, W.~D., {Sagawa}, H., \& {Yoshida}, S. 2012, \prl, 108, 052501

\bibitem[{{M{\"o}ller} {et~al.}(1995){M{\"o}ller}, {Nix}, {Myers}, \& {Swiatecki}}]{1995ADNDT..59..185M}
{M{\"o}ller}, P., {Nix}, J.~R., {Myers}, W.~D., \& {Swiatecki}, W.~J. 1995, Atomic Data and Nuclear Data Tables, 59, 185

\bibitem[{{M{\"o}ller} {et~al.}(2015){M{\"o}ller}, {Sierk}, {Ichikawa}, {Iwamoto}, \& {Mumpower}}]{2015PhRvC..91b4310M}
{M{\"o}ller}, P., {Sierk}, A.~J., {Ichikawa}, T., {Iwamoto}, A., \& {Mumpower}, M. 2015, \prc, 91, 024310

\bibitem[{{M{\"o}ller} {et~al.}(2016){M{\"o}ller}, {Sierk}, {Ichikawa}, \& {Sagawa}}]{2016ADNDT.109....1M}
{M{\"o}ller}, P., {Sierk}, A.~J., {Ichikawa}, T., \& {Sagawa}, H. 2016, Atomic Data and Nuclear Data Tables, 109, 1

\bibitem[{{Mumpower} {et~al.}(2012){Mumpower}, {McLaughlin}, \& {Surman}}]{2012PhRvC..86c5803M}
{Mumpower}, M.~R., {McLaughlin}, G.~C., \& {Surman}, R. 2012, \prc, 86, 035803

\bibitem[{{Mumpower} {et~al.}(2015){Mumpower}, {Surman}, {Fang}, {Beard}, {M{\"o}ller}, {Kawano}, \& {Aprahamian}}]{2015PhRvC..92c5807M}
{Mumpower}, M.~R., {Surman}, R., {Fang}, D.~L., {et~al.} 2015, \prc, 92, 035807

\bibitem[{{Mumpower} {et~al.}(2016){Mumpower}, {Surman}, {McLaughlin}, \& {Aprahamian}}]{2016PrPNP..86...86M}
{Mumpower}, M.~R., {Surman}, R., {McLaughlin}, G.~C., \& {Aprahamian}, A. 2016, Progress in Particle and Nuclear Physics, 86, 86

\bibitem[{{Nedora} {et~al.}(2021){Nedora}, {Bernuzzi}, {Radice}, {Daszuta}, {Endrizzi}, {Perego}, {Prakash}, {Safarzadeh}, {Schianchi}, \& {Logoteta}}]{2021ApJ...906...98N}
{Nedora}, V., {Bernuzzi}, S., {Radice}, D., {et~al.} 2021, \apj, 906, 98

\bibitem[{{Niu} {et~al.}(2009){Niu}, {Sun}, \& {Meng}}]{2009PhRvC..80f5806N}
{Niu}, Z., {Sun}, B., \& {Meng}, J. 2009, \prc, 80, 065806

\bibitem[{{Placco} {et~al.}(2017){Placco}, {Holmbeck}, {Frebel}, {Beers}, {Surman}, {Ji}, {Ezzeddine}, {Points}, {Kaleida}, {Hansen}, {Sakari}, \& {Casey}}]{2017ApJ...844...18P}
{Placco}, V.~M., {Holmbeck}, E.~M., {Frebel}, A., {et~al.} 2017, \apj, 844, 18

\bibitem[{{Radice} {et~al.}(2018){Radice}, {Perego}, {Hotokezaka}, {Fromm}, {Bernuzzi}, \& {Roberts}}]{2018ApJ...869..130R}
{Radice}, D., {Perego}, A., {Hotokezaka}, K., {et~al.} 2018, \apj, 869, 130

\bibitem[{{Roederer} {et~al.}(2024){Roederer}, {Beers}, {Hattori}, {Placco}, {Hansen}, {Ezzeddine}, {Frebel}, {Holmbeck}, \& {Sakari}}]{2024ApJ...971..158R}
{Roederer}, I.~U., {Beers}, T.~C., {Hattori}, K., {et~al.} 2024, \apj, 971, 158

\bibitem[{{Sneden} {et~al.}(2008){Sneden}, {Cowan}, \& {Gallino}}]{2008ARA&A..46..241S}
{Sneden}, C., {Cowan}, J.~J., \& {Gallino}, R. 2008, \araa, 46, 241

\bibitem[{{Surman} \& {Engel}(2001)}]{2001PhRvC..64c5801S}
{Surman}, R. \& {Engel}, J. 2001, \prc, 64, 035801

\bibitem[{{Symbalisty} \& {Schramm}(1982)}]{1982ApL....22..143S}
{Symbalisty}, E. \& {Schramm}, D.~N. 1982, \aplett, 22, 143

\bibitem[{{Vassh} {et~al.}(2024){Vassh}, {Wang}, {Larivi{\`e}re}, {Sprouse}, {Mumpower}, {Surman}, {Liu}, {McLaughlin}, {Denissenkov}, \& {Herwig}}]{2024PhRvL.132e2701V}
{Vassh}, N., {Wang}, X., {Larivi{\`e}re}, M., {et~al.} 2024, \prl, 132, 052701

\bibitem[{{Wang} {et~al.}(2017){Wang}, {Audi}, {Kondev}, {Huang}, {Naimi}, \& {Xu}}]{2017ChPhC..41c0003W}
{Wang}, M., {Audi}, G., {Kondev}, F.~G., {et~al.} 2017, Chinese Physics C, 41, 030003

\bibitem[{{Wang} {et~al.}(2021){Wang}, {Huang}, {Kondev}, {Audi}, \& {Naimi}}]{2021ChPhC..45c0003W}
{Wang}, M., {Huang}, W.~J., {Kondev}, F.~G., {Audi}, G., \& {Naimi}, S. 2021, Chinese Physics C, 45, 030003

\bibitem[{{Wang} {et~al.}(2010){Wang}, {Liang}, {Liu}, \& {Wu}}]{2010PhRvC..82d4304W}
{Wang}, N., {Liang}, Z., {Liu}, M., \& {Wu}, X. 2010, \prc, 82, 044304

\bibitem[{{Wang} {et~al.}(2014){Wang}, {Liu}, {Wu}, \& {Meng}}]{2014PhLB..734..215W}
{Wang}, N., {Liu}, M., {Wu}, X., \& {Meng}, J. 2014, Physics Letters B, 734, 215

\bibitem[{{Wang} {et~al.}(2020){Wang}, {N3AS Collaboration}, {Vassh}, {FIRE Collaboration}, {Sprouse}, {Mumpower}, {Vogt}, {Randrup}, \& {Surman}}]{2020ApJ...903L...3W}
{Wang}, X., {N3AS Collaboration}, {Vassh}, N., {et~al.} 2020, \apjl, 903, L3

\bibitem[{{Watson} {et~al.}(2019){Watson}, {Hansen}, {Selsing}, {Koch}, {Malesani}, {Andersen}, {Fynbo}, {Arcones}, {Bauswein}, {Covino}, {Grado}, {Heintz}, {Hunt}, {Kouveliotou}, {Leloudas}, {Levan}, {Mazzali}, \& {Pian}}]{2019Natur.574..497W}
{Watson}, D., {Hansen}, C.~J., {Selsing}, J., {et~al.} 2019, \nat, 574, 497

\bibitem[{{Wu} {et~al.}(2022){Wu}, {Zhao}, {Zhang}, \& {Meng}}]{2022ApJ...941..152W}
{Wu}, X.~H., {Zhao}, P.~W., {Zhang}, S.~Q., \& {Meng}, J. 2022, \apj, 941, 152

\bibitem[{{Yong} {et~al.}(2021){Yong}, {Kobayashi}, {Da Costa}, {Bessell}, {Chiti}, {Frebel}, {Lind}, {Mackey}, {Nordlander}, {Asplund}, {Casey}, {Marino}, {Murphy}, \& {Schmidt}}]{2021Natur.595..223Y}
{Yong}, D., {Kobayashi}, C., {Da Costa}, G.~S., {et~al.} 2021, \nat, 595, 223

\bibitem[{{Zhu} {et~al.}(2021){Zhu}, {Lund}, {Barnes}, {Sprouse}, {Vassh}, {McLaughlin}, {Mumpower}, \& {Surman}}]{2021ApJ...906...94Z}
{Zhu}, Y.~L., {Lund}, K.~A., {Barnes}, J., {et~al.} 2021, \apj, 906, 94

\end{thebibliography}

\end{document}